\documentclass[aps,reprint,superscriptaddress,showpacs,amsmath]{revtex4-2}

\usepackage{lmodern}

\usepackage{lmodern}
\usepackage[T1]{fontenc}
\usepackage{color}
\usepackage{amstext}
\usepackage{amssymb}
\usepackage{amsmath}
\usepackage{graphicx}
\usepackage{esint}
\newcommand{\etal}{\textit{et al.}}

\begin{document}
\title{Probing hundreds of individual quantum defects in polycrystalline and amorphous alumina}

\author{Chih-Chiao Hung} 
\affiliation{Laboratory for Physical Sciences, 8050 Greenmead Drive, College Park, Maryland 20740, USA}
\affiliation{Quantum Materials Center, University of Maryland, College Park, Maryland 20742, USA} 
\affiliation{Department of Physics, University of Maryland, College Park, Maryland 20742, USA} 
\author{Liuqi Yu}
\affiliation{Laboratory for Physical Sciences, 8050 Greenmead Drive, College Park, Maryland 20740, USA}
\affiliation{Quantum Materials Center, University of Maryland, College Park, Maryland 20742, USA} 
\author{Neda Foroozani} 
\affiliation{Laboratory for Physical Sciences, 8050 Greenmead Drive, College Park, Maryland 20740, USA}
\affiliation{Quantum Materials Center, University of Maryland, College Park, Maryland 20742, USA} 
\author{Stefan Fritz} 
\affiliation{Laboratory for Electron Microscopy, Karlsruhe Institute of Technology, Karlsruhe, 76131, Germany} 
\author{Dagmar Gerthsen} 
\affiliation{Laboratory for Electron Microscopy, Karlsruhe Institute of Technology, Karlsruhe, 76131, Germany} 
\author{Kevin D. Osborn} 
\affiliation{Laboratory for Physical Sciences, 8050 Greenmead Drive, College Park, Maryland 20740, USA}
\affiliation{Quantum Materials Center, University of Maryland, College Park, Maryland 20742, USA}
\affiliation{Joint Quantum Institute, University of Maryland, College Park, MD 20742, USA}
\date{\today}

\begin{abstract}
Quantum two-level systems (TLSs) are present in the materials of qubits and are considered defects because they limit qubit coherence. 
For superconducting qubits, the quintessential Josephson junction barrier is made of amorphous alumina, which hosts TLSs.
However, TLSs are not understood generally -- either structurally or in atomic composition. 
In this study, we greatly extend the quantitative data available on TLSs by reporting on the physical dipole moment in two alumina types: polycrystalline $\mathrm{\mathrm{\gamma-Al}_{2}\mathrm{O}_{3}}$
and amorphous $\mathrm{a-Al}\mathrm{O_{x}}$. 
To obtain the dipole moments $p_z$, rather from the less-structural coupling parameter g, we tune individual TLSs with an external electric field to extract the $p_z$ of the TLSs in a cavity QED system.
We find a clear difference in the dipole moment distribution from the film types, indicating a difference in TLS structures. 
A large sample of approximately 400 individual TLSs are analyzed from the polycrystalline film type. 
Their dipoles along the growth direction $p_z$ have a mean value of 2.6$\pm$0.3 Debye (D) (0.54$\pm$0.06 e$\mathring{A}$) and standard deviation $\sigma$ = 1.6$\pm$0.2 D (0.33$\pm$0.03 e$\mathring{A}$). 
The material distribution fits well to a single Gaussian function. 
Approximately 200 individual TLSs are analyzed from amorphous films. Both the mean $p_z$ =4.6$\pm$0.5 D (0.96$\pm0.1$ e$\mathring{A} $) and $\sigma$ =2.5$\pm$0.3 D (0.52$\pm0.05$ e$\mathring{A}$) are larger. Amorphous alumina also has very large $p_z$, > 8.6 D (1.8 e$\mathring{A} $), in contrast to polycrystalline which has none of this moment. These large moments agree only with oxygen-based TLS models. Based on data and the candidate models (delocalized O and hydrogen-based TLSs), we find polycrystalline alumina has smaller ratio of O-based to H-based TLS than amorphous alumina.
\end{abstract}

\maketitle
\section{Introduction}

Long coherence times are essential for quantum information processing
and this implies high-quality
Josephson junctions (JJs) in superconducting qubits \cite{firstqubit,qubitcitation2,fluxonium,qubitcitation3,Cshuntflux,supremacy}.
For over a dozen years, quantum tunneling two-level systems (TLSs) have been
known to be defects that cause loss and limit coherence of qubits \cite{TLS-qubitcouple2004,JMartinislossSiO2SiNxphase6D}.
In addition, TLSs create telegraphic noise and 1/f noise \cite{TLSnoiseresonatorJ.Burnett,TLSowntelegraphnoise,benchmarkingT1drop}
in superconducting qubits \cite{TLSnoiseonqubit,fluctuationT1dropJM,TLSnoiseonqubit2},
semiconducting qubits \cite{Johnnichol}, and astronomy photon detectors
\cite{GaoTiNphotondetector,Gaothesis}. There are several strategies
to improve the qubit coherence time such as material optimization \cite{dielectriclossdifferentmaterail,LockTiN,PappasTiN,dielectriclossandclean},
surface treatments \cite{dielectriclossandclean,imperfectinterface,nativealox},
and engineering of the qubit geometry to decrease the participation of
TLSs \cite{3Dcavity,yale3dqubitgeo,GeochangeIBM}.

In the quintessential JJ, an amorphous alumina barrier is 
grown thermally on the surface of aluminum \cite{firstqubit,qubitcitation2,qubitcitation3}.
Loss tangents of amorphous alumina in JJs \cite{dielectriclossdifferentmaterail}
and in grown films \cite{CDengthesis} are measured to be approximately $2\times10^{-3}$,
much higher than that in crystalline alumina from the sapphire substrates used for qubit fabrication \cite{Sapphirelowloss}. Accordingly, amorphous materials are believed to have higher loss than crystalline ones due to additional tunneling degrees of freedom (TLSs) in the former. HBN and
other 2D materials are being investigated for JJs \cite{MoS2JJ,hBNJJ,Charles2Dmaterial},
but an alternative method to improve the JJ barrier uses
annealed crystalline alumina \cite{crystalAl2O3phaseQ,crystalAl2O3transmon}.
Crystalline alumina studies show a decrease in both TLS density \cite{crystalAl2O3phaseQ}
and TLS-qubit couplings, $g$, relative to amorphous alumina \cite{crystalAl2O3transmon}.

Recent TLS analysis techniques use dc-tuned electric \cite{BahmanDCbiasTLS,LisenfieldalmuniaparallelC,bilmes2020vbias,Graaf2021vbias}
or strain field \cite{Straintune2010,StraintuneTLS2012,straintune4transmissionline}
for the observation of individual nanoscale defects. According to the Standard Tunneling Model (STM) \cite{STM1,STM2}, individual TLSs have a dipole moment $p$, transition energy $E$ and
tunneling energy $\Delta_{0}$. Generally, the TLS is described as tunneling charge presumed to
be an atom or small group of atoms, though a recent study reports on a possibility of trapped quasiparticles \cite{Graafquasipotential2020}. 
Their identification is a 50 years old mystery \cite{yu2021two}.

TLSs have an ac-coupling to quantum systems,
\begin{equation}
g=\frac{\Delta_{0}}{E}\,\frac{2\,p_{z}\mathrm{E}_{rms}}{\hbar},
\end{equation}
which is related to dipole moment component $p_z$ along zero-point
fluctuation of electric field $\mathrm{\mathbf{E}}_{rms}$.
However, the most common measurement of $g$ does not allow extraction of $p_z$ because $\Delta_0$ is unknown or $\mathrm{E}_{rms}$ is not uniform \cite{JMartinislossSiO2SiNxphase6D,crystalAl2O3phaseQ,crystalAl2O3transmon}.
On the other hand, static dc-tuned measurements allow measurements of individual $p_{z}$ \cite{BahmanDCbiasTLS,LisenfieldalmuniaparallelC} and dynamically biased experiments extract averaged $p_{z}$ \cite{MoeACEfield}.
Such dynamical bias can induce Landau-Zener transitions, and recent work shows that a resonator can even exhibit dynamical decoupling using these transitions \cite{Dynamicaldecoupling}.
Recently, voltage bias gates are added above or below the target area to tune TLSs \cite{bilmes2020vbias,Graaf2021vbias}. 
The extractions of dipoles are possible from $g$ and extracted position, but the distribution is given as a function of $g$ rather than $p_z$ \cite{bilmes2020vbias}.
Also, only small samples of $p_z$ were directly extracted in the past: 13 in amorphous alumina \cite{LisenfieldalmuniaparallelC} and 64 in silicon nitride \cite{BahmanDCbiasTLS}.
To the best of our knowledge, a comparison of $p_{z}$ in two different materials has not been performed in a single study.

In this letter we study individual TLSs in both
polycrystalline alumina $\mathrm{\gamma-Al\mathrm{_{2}O}_{3}}$ and amorphous alumina $\mathrm{a-Al}\mathrm{O_{x}}$.
We follow the circuit schematic of Ref. \cite{BahmanDCbiasTLS}, and we now name it an Electrical-Bridge Quantum-Defect Sensor (EBQuDS).
The TLSs were analyzed in films with an approximate thickness of 20 nm. The TLSs in $\mathrm{\gamma-Al\mathrm{_{2}O}_{3}}$ films
are relatively stable, and allow us to obtain a large distribution
of 394 TLS dipole moments $p_{z}$. In $\mathrm{a-Al}\mathrm{O_{x}}$
films, 189 TLS moments are extracted despite higher TLS noise. 
Compared to the $\mathrm{\gamma-Al\mathrm{_{2}O}_{3}}$ film, larger averaged dipole and standard deviation $\sigma$ are extracted in $\mathrm{a-Al}\mathrm{O_{x}}$ films and 10\% of TLSs have larger dipole than the maximum extracted in $\mathrm{\gamma-Al\mathrm{_{2}O}_{3}}$.
Specific TLS structures were proposed using available information from recent work on density functional theory (DFT) analysis of TLSs in alumina, where both hydrogen \cite{AlOxmodel,AlOxmodel2}
and oxygen \cite{AlOxmodel3,AlOxmodel4similarto3,AlOxmodel5}
based TLS have been proposed as the interstitial defects.
A comparison of the dipole moments for both $\mathrm{\gamma-Al\mathrm{_{2}O}_{3}}$
and $\mathrm{a-Al}\mathrm{O_{x}}$ film types allows possible TLS origin identification.

\section{Method}

\begin{figure}
\begin{centering}
\includegraphics[width=8.6cm]{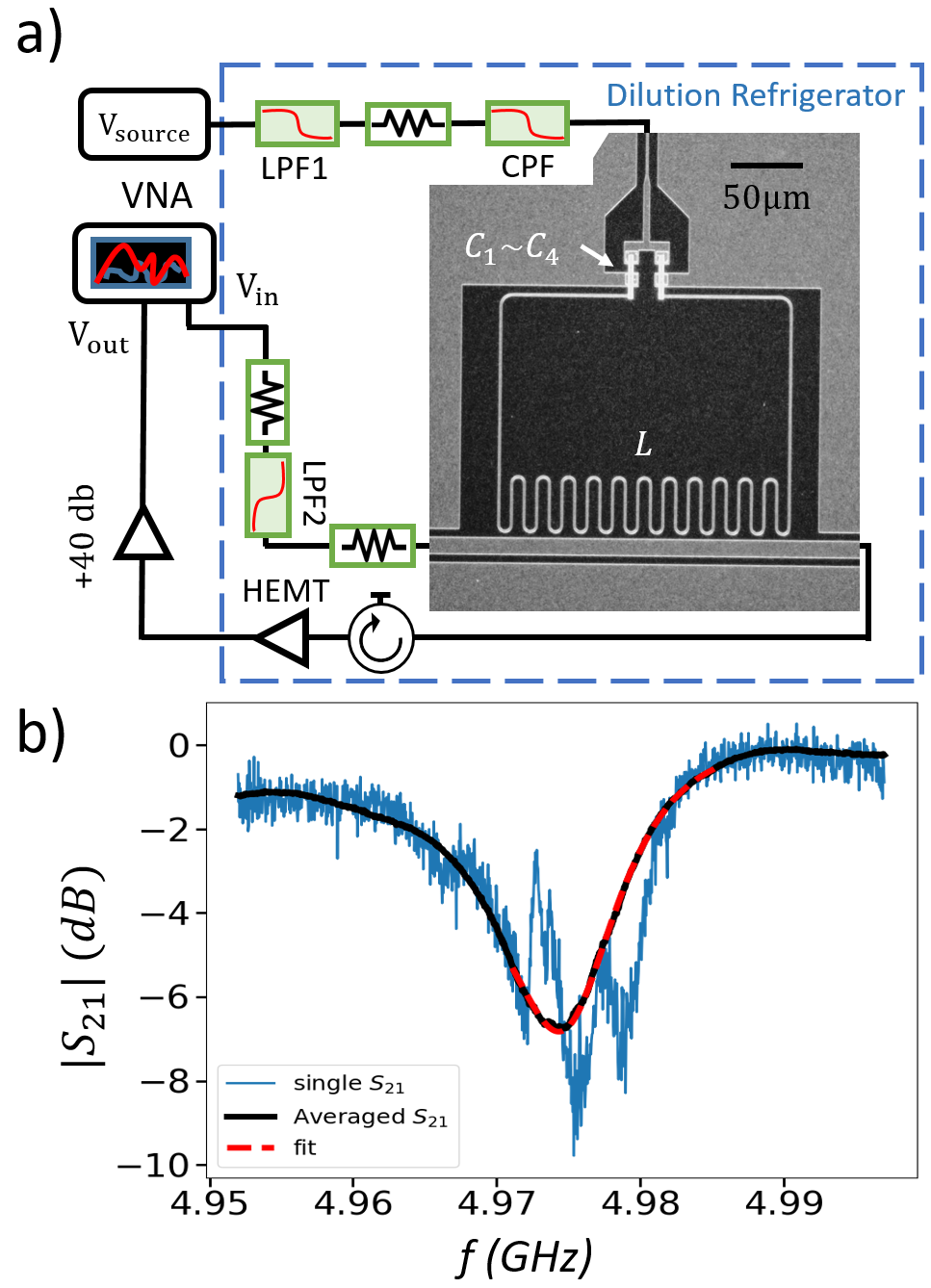} 
\par\end{centering}
\caption{(a) Optical image of microwave resonator with an abbreviated wiring
schematic. The source voltage is filtered through an RC low pass filter
and a copper powder filter (CPF). The combination of resistance in
the RC filter and the 3dB attenuator gives $\mathrm{V_{bias}}$ = 9.56 $\times \mathrm{10^{-3}\times V_{source}}$.
(b) $\left|S_{21}\right|$ of one $\mathrm{\gamma-Al_{2}\mathrm{O}_{3}}$
resonator. It shows multiple dips indicating TLSs strongly coupled
to the resonator. The black curve is the ensemble average
$\left|S_{21,avg}\right|$, obtained by averaging $S_{21}$ from different bias voltages. The
intrinsic quality $\mathrm{Q_{i}}$ is $\mathrm{1\,/\,(1.47\times10^{-3})\approx\,680}$
according to the red fitting line .\label{fig1 optical and layout}}
\end{figure}

Fabrication starts by $\mathit{in\:situ}$ growth of Al/alumina/Al
trilayers on a 3-inch Si substrate, where the alumina is the material
hosting the TLSs. The $\mathit{in\:situ}$ method prevents substantial hydrogen contamination, but diffusion of hydrogen is also difficult to prevent in standard lithographic processing \cite{van2003universal}. Then, a first BCl$_{3}$ etch forms a mesa into
the top 2 layers, defining 4 equal capacitors ($\mathrm{C_{1}-C_{4}}$).
Next, a second BCl$_{3}$ etch forms the base-metal including a resonator
inductor L and ground plane. Finally, silicon nitride is deposited
as a wiring dielectric, vias are etched by SF$_{6}$, and an Al wiring
layer is defined to connect the inductor to the capacitors. This
creates the final resonator structure as shown in Fig. \ref{fig1 optical and layout} (a).
Alumina in the dielectric layer is designed to have an
approximate thickness $d$ = 20 nm and a volume $\mathbb{V}=1.11\times10^{-17}m^{3}$
in each capacitor. 

The two sample types are grown in separate chambers.
The polycrystalline sample has 20 nm thick $\gamma-\mathrm{Al_{2}O_{3}}$ film deposited by electron-beam evaporation from 99.99\% purity $\mathrm{Al_{2}O_{3}}$ pellets with a base pressure of < $5\times10^{-7}$ torr and alumina is sandwiched by two 100 nm thick aluminum films.
On the other hand, the $\mathrm{a-Al}\mathrm{O_{x}}$ film is 14.7 nm thick with x = 1.3 $\pm$ 0.1 and is grown by iterating eight rounds of 1 nm Al deposition followed by static oxidation at 9.5 mbar of oxygen at 250 $^{\circ}$ C as described in ref. \cite{Stefan1}.
Given this oxidation condition, no long-range
ordered (crystalline) structure was detected from transmission electron
microscopy and only 3\% unoxidized aluminum was found inside the $\mathrm{a-Al}\mathrm{O_{x}}$
layer. Transmission electron microscopy shows that the thickness can vary from 10 to 20 nm in some rare cases.

The resonator inductively couples to the transmission line so that
a 2-port microwave transmission measurement can be carried out. The
applied voltage from room temperature is filtered by an RC filter,
3dB attenuator and a copper powder filter. It generates an dc biased field
$\mathrm{E_{ex}}$ across each capacitor. The maximum $\mathrm{E_{ex}}$
is $90\mathrm{\,kV/m}$ with which we observe no refrigerator heating,
thus no significant leakage current. Two resonators
were fabricated per chip with nominally the same capacitors, but with
different value inductors, giving resonance frequencies of approximately
$f_{0}=$ 5.0 GHz and 4.4 GHz. The resonators were measured at or
below 60 mK. A less than 1 probing photon number $\bar{n}$ is used
for all the reported data, to allow observation of TLSs near their
ground state.

With a known external field $\mathbf{\mathrm{E_{ex}}}$,
the asymmetry energy $\Delta$ is shifted as $\Delta'\,=\,\Delta+2\,p_{z}\,\mathbf{\mathrm{E_{ex}}}$ \cite{BahmanDCbiasTLS}.
Therefore, the resultant TLS energy is 
\begin{equation}
E\,=\,\sqrt{(\Delta+2\,p_{z}\,\mathbf{\mathrm{E_{ex}}})^{2}+\Delta_{0}^{2}}\label{eq:hyperbola}
\end{equation}
The resonator constitutes a circuit QED system with a Jaynes-Cummings
model modified for many TLSs. TLSs can be resolved individually by
the resonator when the cooperativity $g^{2}/\gamma_{\mathrm{TLS}}\,\kappa\,\geq 1$ \cite{nontunedTLSbahman}, where $\gamma_{\mathrm{TLS}}$
is the TLS decay rate for the strongly coupled TLS, and $\kappa=\kappa_{e}+\kappa_{i}$
is the resonator decay rate from external coupling and internal loss.
We increase the $g$ and cooperativity by reducing $\mathbb{V}$, since $\mathrm{E}_{rms}=\sqrt{hf_{0}/8\epsilon_{r}\epsilon_{0}\mathbb{V}}$
in our parallel-plate capacitor resonator \cite{JMartinislossSiO2SiNxphase6D}.

A single transmission trace $\left|S_{21}\right|$ is shown in Fig. \ref{fig1 optical and layout} (b)
from a $\mathrm{\gamma-Al_{2}\mathrm{O}_{3}}$ resonator. Within the
bandwidth of the resonator, a few fine resonance dips reveal the energies
of individual TLSs. However, the TLSs often only couple weakly to
the resonator. Therefore, we obtain an approximate intrinsic material loss tangent
$\tan\delta_{0}$ by using the averaged $S_{21}$ traces from
different voltage biases, yielding an ensemble-averaged $S_{21,avg}$.
The |$S_{21,avg}$| component of the result is shown as the solid
black curve in Fig. \ref{fig1 optical and layout} (b). A fit (dashed
red curve) to $S_{21,avg}$ yields $\tan\delta_{0}\,=\,1/Q_i\,=\,1/680\,=\,1.5\times10^{-3}$ and the external (or coupling)
quality factor is extracted as $\mathrm{Q_{e}}\,=2\pi f_{0}/\kappa_{e}=\,590$.
The same procedure performed on $\mathrm{a-Al}\mathrm{O_{x}}$ gives
an intrinsic loss tangent of $\tan\delta_{0}\,=\,1/\overline{Q_i}\,=\,1/1020\,=\,9.8\times10^{-4}$,
which is smaller than that of $\mathrm{\gamma-Al_{2}\mathrm{O}_{3}}$. $\overline{Q_i}$ is the averaged $Q_i$ of $\mathrm{a-Al}\mathrm{O_{x}}$ and its $Q_i$ are distinct with every cooldown unlike $\gamma-\mathrm{Al_{2}O_{3}}$ (Appendix \ref{aAlOxfit}). 
Our loss tangents are similar to alumina with different growth methods, where the loss is measured at $\mathrm{tan\delta_{0}\,}=\,1.6\times10^{-3}$ \cite{JMartinislossSiO2SiNxphase6D,JJdefectmeter}, $7\times10^{-4}$
\cite{MoeaAl2O3}, and $1.6\times10^{-3}$
\cite{pappas2011two}. Below we discuss TLSs measured in both alumina
film types using two resonators for each type.

\section{Results and Discussion}

\begin{figure}[t]
\centering{}\includegraphics[width=8.6cm]{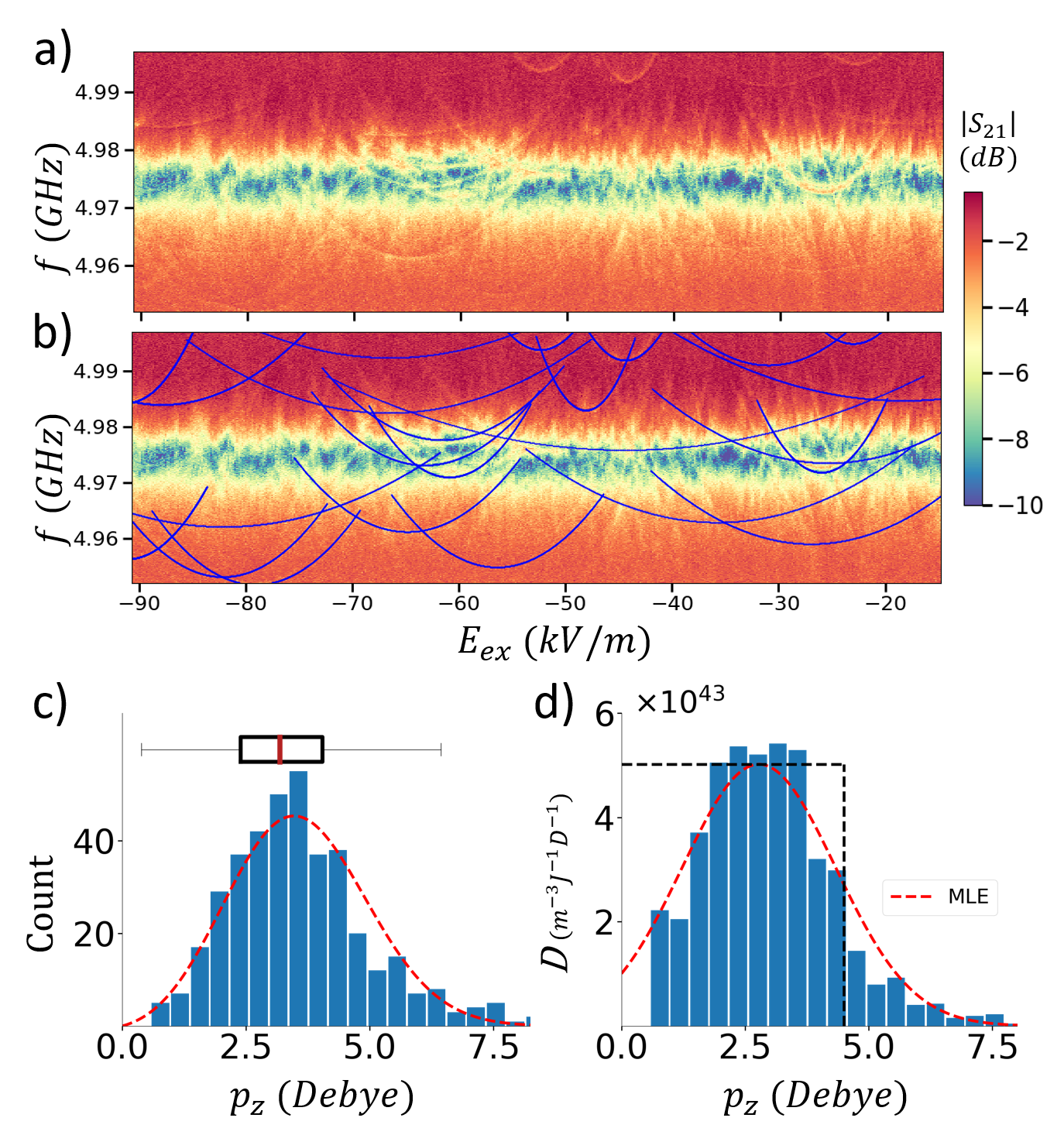}\caption{Data on nanoscale thick $\gamma-\mathrm{Al_{2}O_{3}}$(polycrystalline) film
in cQED system (a) Color scale plot of transmission $|S_{21}|$ vs.
frequency $f$ and electric field $\mathrm{E_{ex}}$. Data show a main resonance at 4.974 GHz. Several local minima in $|S_{21}|$
reveal the energy of individual TLSs.
(b) Several TLSs in blue hyperbolas are fitted to the energy model (Eq. \ref{eq:hyperbola}), where $p_{z}$ comes from the curvature of the energy hyperbola.
(c) The entire data set from this film type yields 394 moments $p_{z}$.
The red dashed line is a Gaussian function multiplied
by $p_{z}$ as fit to data. A Seaborn box is plotted, where the red color line is the mean line and the right and left side of box represent 25th and 75th percentage of dipole data.  
(d) Material TLS density of $p_{z}$ after accounting
for the experimental weighing factor. A red dashed Gaussian line is shown with the corresponding fit parameters
in (c). From the fit we report the material mean $p_{z}$ of 2.6 D and standard deviation $\sigma$ =  1.6 D.
The black dashed line illustrates a possible material density if we assume an isotropic TLS
direction.\label{Fig2 sweep dc and fit}}
\end{figure}

Fig. \ref{Fig2 sweep dc and fit} (a) shows a TLS spectrum example,
$\left|S_{21}\right|$, as a function of frequency $f$ and the dc-field
$\mathrm{E_{ex}}$ for $\mathrm{\gamma-Al_{2}\mathrm{O}_{3}}$ measured
on one of the two resonators during one cool-down. TLS energies exhibit
hyperbolic energies versus $\mathrm{E_{ex}}$, as shown in Fig. \ref{Fig2 sweep dc and fit}
(a) in agreement with Eq. \ref{eq:hyperbola}. Similar spectra have
been observed and analyzed in a previous study on silicon nitride
\cite{BahmanDCbiasTLS}. One can estimate the $\Delta_{0}$ from
the minimum of the TLS energy ($hf_{m}$$=\Delta_{0}$), and $p_{z}$
from the hyperbola -- a steeper curvature gives a larger
dipole moment. A optimized Monte Carlo fit is performed on each TLS
energy to extract $p_{z}$ of the specific TLS (see Appendix. \ref{Montefit}). Only well defined
TLS energy curves are selected for analysis. Example fits are
plotted as blue hyperbolas in Fig. \ref{Fig2 sweep dc and fit} (b).

TLSs change their energies randomly during cool-downs from room temperature.
From four different cool-downs, we created different sets of TLSs in the two
resonators with one material type. 
According to the standard TLS distribution \cite{STM1}, TLSs have log-uniform tunneling energies such that there are negligible distribution changes for TLS tunneling energies that are only 0.6 GHz different in our two resonator frequencies (see Appendix. \ref{tworesHis}).
We therefore combine all the data from different runs in the two resonators to
enlarge the sampling number and improve the statistics: a total of
394 TLSs from two resonators are analyzed to form the measured $p_{z}$
distribution $H(p_{z})$ with an average of 3.5 $\pm0.4$
Debye (D) shown in Fig. \ref{Fig2 sweep dc and fit} (c). The accuracy of extraction is limited by the uniformity of thickness of alumina rather than fitting process. A large amounts of individual TLSs allow relatively accurate representation
of the TLS moments in $\gamma-\mathrm{Al_{2}O_{3}}$.

Though the measured polycrystalline distribution $H(p_{z})$ has a mean value of
3.5 D, this is not an intrinsic material property. At a
given electric field bias range $\Delta\mathrm{E}_{ex}$, TLSs with
larger dipole moments have a larger shift in asymmetry energy $\Delta'$
relative to smaller moments, and this leads to a higher probability
of the former moments having their energy minimum within the resonator
bandwidth (see Appendix \ref{HtoD}
and Ref \cite{BahmanDCbiasTLS}). The intrinsic material TLS dipole
distribution $D(p_{z})$ is related to TLS material density
$P_{0}\,=\,\intop D(p_{z})\,dp_{z}$ (in units of $J^{-1}m^{-3}$),
and can be calculated from
$D(p_{z})\,=\,\frac{1}{\mathbb{V}}\,\frac{H(p_{z})}{2\,p_{z}}\,\frac{1}{\,\Delta E_{ex}}\,\frac{f_{0}}{\Delta f_{0}}$
, where $f_{0}$ is the resonator frequency and
$\Delta f_{0}$ is the frequency span of the $S_{21}$ measurement.
The red dashed line in Fig. \ref{Fig2 sweep dc and fit} (c) shows a fit using a maximum likelihood estimation (MLE) method. The fitting function is a modified Gaussian distribution,
which is a Gaussian distribution multiplied by $p_{z}$. The TLS material density $D(p_{z})$
is shown in Fig. \ref{Fig2 sweep dc and fit} (d), where the red line
shows a Gaussian function matching the fit parameters
in Fig. \ref{Fig2 sweep dc and fit} (c). 

From the fit, we find that the polycrystalline material distribution
$D(p_{z})$ has a fit mean dipole moment of $\bar{p}_{z}$=
2.6 $\pm0.3$ D (=0.54$\pm0.05$ e$\mathring{A} $) and $\sigma$ =
1.6 D (=0.33 e$\mathring{A} $). The computed TLS density $P_{0}$ is $1.0 \pm0.1\times10^{44}\mathrm{(J^{-1}m^{-3})}$.
This computed value of TLS density along with the dipole moments agrees
with the measured loss tangent. As a result we used the material units
in panel (d) for the distribution $D(p_{z})$. 

For amorphous samples
or random voids within polycrystals, we expect TLS dipoles
to be random in angle (isotropic). 
For a case of
one single dipole magnitude $p_{0}$ and uniform distribution in  $\mathrm{cos}\theta$, where $\theta$ is the angle of dipole to z-axis, results isotropy.
Therefore, $D(p_{z})$ is expected to be independent of $p_{z}$ until the maximum value $p_{0}$.
As a guide to the eye, an isotropic distribution (random
angle) with dipole moment $p_{0}$ = 4.5 D is shown
as a black dashed line in Fig. \ref{Fig2 sweep dc and fit} (d). The positive slope in the observed distribution 
indicates that we have a departure from isotropic distribution (isotropic TLSs give only non-positive slope). Thus, data in Fig. \ref{Fig2 sweep dc and fit}
(d), shows that $\gamma-\mathrm{Al_{2}O_{3}}$ TLSs can be different
than the standard model for TLSs (designated for amorphous samples \cite{STM1,STM2}).
The anisotropic angular distribution may be caused by the polycrystalline
film texture (crystallite orientation) which influences the TLS orientation.

\begin{figure}
\centering{}\includegraphics[width=8.6cm]{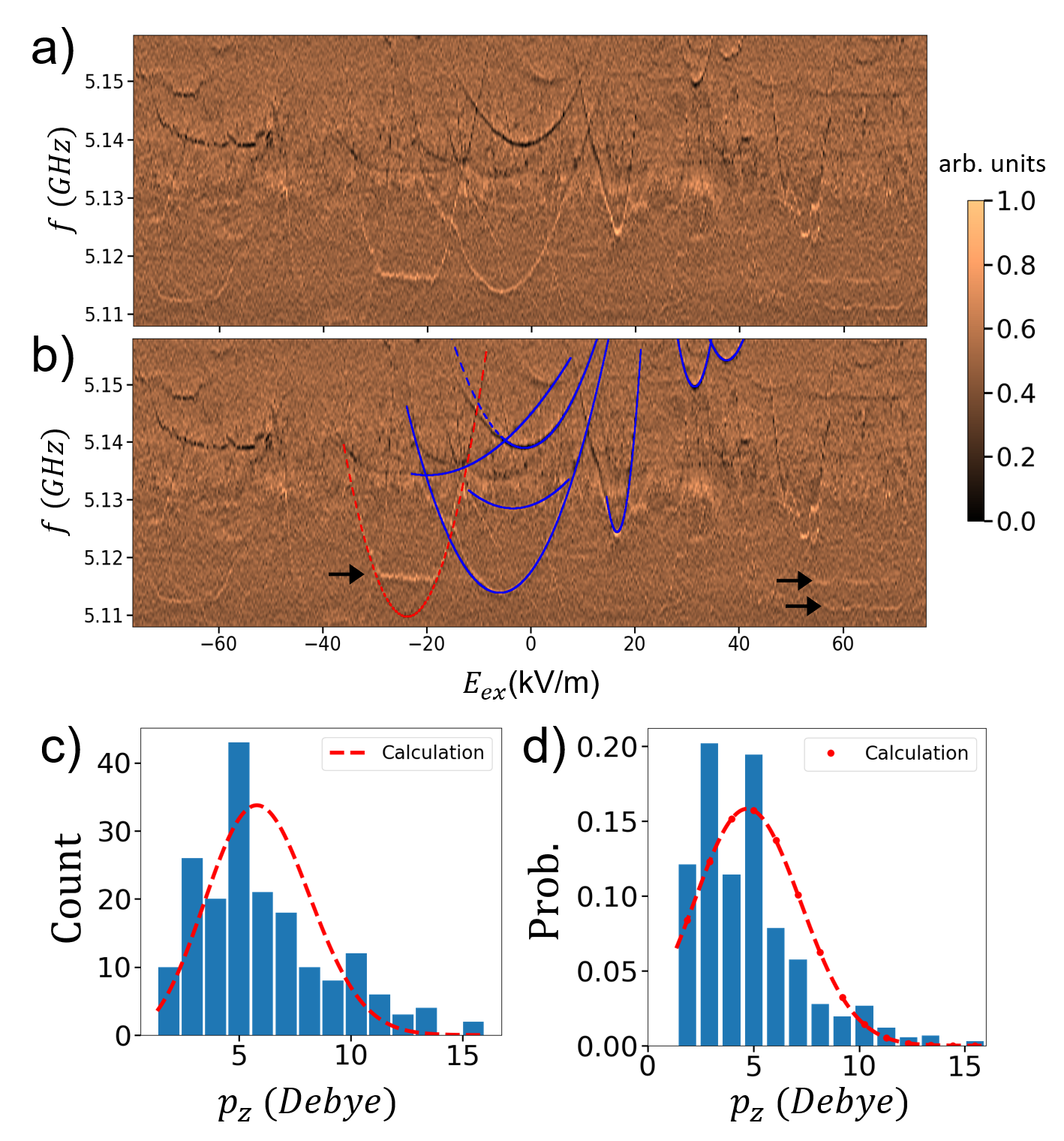}\caption{(a) Example of processed transmission data on amorphous $\mathrm{a-Al}\mathrm{O_{x}}$
film resonator versus frequency $f$ and external electric field $E_{ex}$
(see main text). The LC resonance is approximately 5.132 GHz (no longer visible after  data processing). (b) The same data with traces fit to TLS energy function. 7 fitted
dipole moments are extracted at values of 3.2 - 12.7 D. TLSs whose
energy does not depend on bias voltage are marked by arrows. Curved
red dashed trace shows an anomalous TLS; it switches
between a hyperbola and a nearly constant energy of 5.116 GHz at $E_{ex}=-30$
kV/m and $E_{ex}=-18$ kV/m. 
(c) The measured distribution of 189 TLSs and a Seaborn box. The green line represents the largest dipole measured in $\gamma-\mathrm{Al_{2}O_{3}}$. 
(d) The probability of the material TLS dipole distribution. Dashed
lines in (c) and (d) are modified and regular Gaussian functions and obtained by two methods. The red lines are acquired by calculation, and he yellow lines are acquired by fitting. We report a mean dipole = 4.6 $\pm0.5$ D and $\sigma$ = 2.5 $\pm0.3$ D. 
Analysis on missing extracted dipoles was needed in the amorphous film, but it also gives a mean value much larger than in the $\gamma-\mathrm{Al_{2}O_{3}}$ (see main text and Appendix \ref{overestimation}). \label{fig3 :Histogram}}
\end{figure}

Fig. \ref{fig3 :Histogram} (a) shows transmission spectroscopy results
for $\mathrm{a-Al}\mathrm{O_{x}}$. However, the $\mathrm{a-Al}\mathrm{O_{x}}$ spectra are not as clear as in $\gamma-\mathrm{Al_{2}O_{3}}$ due to higher noise in the spectra, despite using the same setup. To improve the TLS signal contrast, the transmission
($S_{21}$) data is shown after processing, unlike the polycrystalline
film data shown earlier. For our first processing step we chose to
subtract $S_{21,avg}$ from $S_{21}$, and increase further contrast 
using the formula $\,(S_{21}(dB)\,-\,S_{21,avg}(dB))\times|S_{21,avg}|$.
As a second and third processing step, we apply a low-pass filter
in the frequency direction and then take the derivative with respect
to frequency. 
The final result is plotted in arbitrary unit (AU) in Fig. \ref{fig3 :Histogram} (a) and (b) and we added hyperbolic fitting traces (blue lines) to the panel (b).

As we will show in detail below, TLSs within $\mathrm{a-Al}\mathrm{O_{x}}$
are less stable than those in $\gamma-\mathrm{Al_{2}O_{3}}$ -- the
former TLSs show sudden switchings in energy or even become invisible in time
within the resonator bandwidth, making it more difficult to identify
individual TLSs. This leads to a higher error in the Monte Carlo
fit. We also observed energy features that are almost
independent of $E_{ex}$, as indicated by black arrows. They are not expected because all coupled TLSs should
be frequency tunable in the device. 
At -8 kV/m, one hyperbola seems to change slope (as indicated by the start of a blue dashed line), although instead this event may represent a transition in observing two separate TLSs.

Surprisingly, one of them seems
to be only partially described by hyperbola in red-dashed curve. 
Increasing bias voltage at -30 kV/m, this TLS seems to switch from a regular TLS state to an unknown state which has constant
transition energy under bias until -19 kV/m, and finally switch back to normal TLS behavior. 
This indicates an unexpected state near its energy minima $\Delta_0$, which we believe has not been identified previously.

Unlike the other sample, the amorphous films show most of the hyperbolas
from TLSs in the bias range of $-30$ to $30$ $\mathrm{kV/m}$ (more data in \cite{supplementary}). Outside
of this range TLSs tracks are seen, but they don't trace out a smooth
hyperbola. Furthermore, we find that the most TLSs do not appear twice after repeating the voltage scanning within the same cool-down.
In a small fraction of TLS hyperbola (< 3\%) a TLS hyperbola has the same dipole and the minimum in energy
is within 1 MHz, such that it is regarded as the same TLS and disregarded
in distribution.

In $a-\mathrm{Al_{2}O_{3}}$, we
identify and analyze a total of 189 TLSs using multiple field sweeps
and cool-downs according to the above procedure. 
The measured distribution $H(p_{z})$ with counts and the probability of material distribution $D(p_z)$ are shown in Fig. \ref{fig3 :Histogram} (c).
The extracted $p_{z}$ shows a broad range in value from 0.5
to 16 D with an average of 6.0 D and the interquartile
range (range from the 25th to 75th points) of 3.8 D. Because
of the large deviation, we cannot get a reasonable fitting to a Gaussian
from the MLE method as shown in yellow lines, which have mean value of 1.6 D and standard deviation of 4.2 D. Instead, we calculate the material average dipole
moment $\bar{p}_{z}$ for $D(p_{z})$ from $H(p_{z})$,
using $\bar{p}_{z}=\intop p_{z}D(p_{z})dp_{z}/\intop D(p_{z})dp_{z}\,=\,$$\intop H(p_{z})dp_{z}/\intop p_{z}^{-1}H(p_{z})dp_{z}$
, and the standard deviation in a similar way. From this we find $\bar{p_{z}}\,=\,4.6 \pm 0.5$ (= 0.96$\pm0.1$ e$\mathring{A} $)
and $\sigma$ = 2.5$\pm$0.3 D (= 0.52$\pm0.05$ e$\mathring{A} $).
A Gaussian curve with these parameters are plotted as red dashed line.

The calculated loss from the dipole distribution is $3.2\times10^{-4}$ which is smaller than
$9.75\times10^{-4}$ reported above.
The missing TLS extraction happens in two ways.
On one hand, larger $p_z$ TLSs have a higher possibility to interact with other TLSs and their frequencies are prone to switch to other states or diffuse. 
On the other hand, small $p_z$ TLSs require longer time to acquire TLS hyperbolas and the signal-to-noise ratio is smaller due to smaller coupling to the resonator. 
As a result, the TLS extraction does not include most of the TLS unlike the extraction in $\gamma-\mathrm{Al_{2}O_{3}}$. 
We estimate the minimum mean moment of amorphous alumina from missing TLSs. We make no claim about anisotropy in this film since the missing TLSs may create one of the peaks in the distribution of $p_z$ (Appendix \ref{overestimation}).

\begin{figure}
\includegraphics[width=8.6cm]{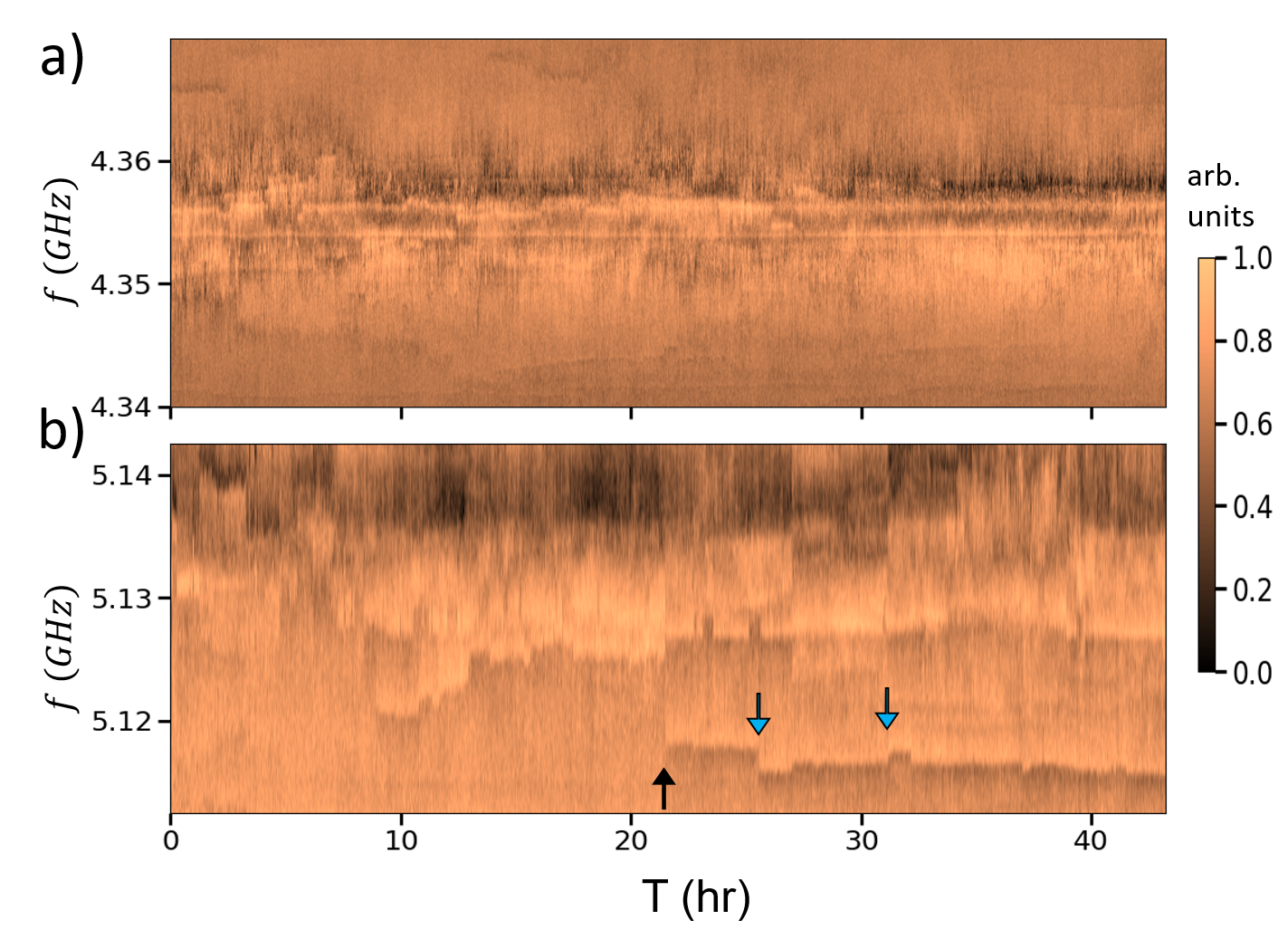}\caption{Time dependence of processed $|S_{21}|$ with a frequency range $\geq$ 30 MHz. (a) $\gamma-\mathrm{Al_{2}O_{3}}$
spectroscopy versus time shows the TLSs are relatively stable in frequency
near the transmission minimum for 10s of hours. (b) $\mathrm{a-AlO_{x}}$
spectroscopy versus time shows relatively large TLS energy switching and drift.
\label{fig4--spectroscopy}}
\end{figure}

To decipher the role of TLS-TLS interaction, we next conduct temporal
spectroscopes for the two different film types: $\gamma-\mathrm{Al_{2}O_{3}}$
and $\mathrm{a-Al}\mathrm{O_{x}}$. Fig. \ref{fig4--spectroscopy}
shows processed $S_{21}$ traces observed over many hours. This resonant TLS noise is believed to be caused by interactions with thermally excited low-frequency TLSs \cite{fluctuationT1dropJM,InteractingTLS1}. As shown in Fig. \ref{fig4--spectroscopy}
(a), TLSs biased at 0 V in $\gamma-\mathrm{Al_{2}O_{3}}$ films near
the resonance frequency are relatively stable -- their energies drift
by less than 2 MHz over tens of hours. On the contrary, TLSs in $\mathrm{a-Al}\mathrm{O_{x}}$
behave similar to Ref. \cite{fluctuationT1dropJM}. TLSs show irregular
drifts of more than 5 MHz, including multiple telegraphic switching
events (blue arrows) and abrupt TLS shifts (black arrow).
Due to the larger dipole moments observed in $\mathrm{a-Al}\mathrm{O_{x}}$,
we expect a larger interaction than that in $\gamma-\mathrm{Al_{2}O_{3}}$
(assuming that the low-frequency thermally activated TLSs are similar
to the high frequency ones). The large unstable behaviors shown in
the amorphous films occur in a few hours, e.g., a 4 D
hyperbola track in $\mathrm{a-Al}\mathrm{O_{x}}$ data took about
5 hours to obtain.

Comparable results on amorphous alumina exist. One study found $p_z$ in the range of 2.3 -7.4 D, using a few analyzed TLSs \cite{straintune4transmissionline}. 
Other field tuned measurements in a-AlOx, studying the barrier of JJs, detected several moments with $p_z$ = 1.0 - 2.9 D \cite{LisenfieldalmuniaparallelC}. 
Measurements of the transition dipole moments in a-AlOx of JJs indicate that $p_z$ $\leq$ 6.0 D \cite{JMartinislossSiO2SiNxphase6D} and $p_z$ $\leq$ 4.8 D
\cite{T1oftlsinjjcouplingG} in two TLS dipoles measured. The existing data on amorphous
alumina TLSs seems consistent with our observations, though the growth methods are slightly different.

Finally, we return to comment on the possible origins of TLSs in our alumina from a comparison to recent TLS DFT \cite{AlOxmodel,AlOxmodel2,AlOxmodel3,AlOxmodel4similarto3} and molecular dynamics \cite{AlOxmodel5} simulations on aluminum oxide. 
Two models related to hydrogen(H) suggest total dipole moment $p$ < 3.0 D \cite{AlOxmodel,AlOxmodel2}. 
Holder \etal\,\,find that hydrogen aluminum-vacancy TLSs $\mathrm{V_{Al}-H}$ in $\alpha-\mathrm{Al_{2}O_{3}}$ have $p$ = 3.0 D \cite{AlOxmodel}. 
Separately, Gordon \etal\,\,simulated the interstitial hydrogen in $\alpha-\mathrm{Al_{2}O_{3}}$ at various two oxygen(O) atoms distances where $p$ =  2.2 – 2.7 D \cite{AlOxmodel2}. 
Besides H-based simulations, two models of O-based TLSs suggest $p$ > 4.2 D. 
DuBois \etal\,\,studied models of delocalized oxygen atoms with six neighboring aluminum atoms \cite{AlOxmodel3,AlOxmodel4similarto3}.
They found oxygen deficient $\mathrm{AlO_{x}}$ for x = 1.25 by varying distances between O and Al atoms, with $p$ =  4.2 - 6.5 D for TLSs with tunneling energy $\Delta_0/h$ = 4 GHz \cite{AlOxmodel3}.
Additionally, Paz \etal\,\,find natural bi-stable structures in amorphous alumina including only Al and O atoms and calculate an average $p$ =  4.2 D from 7 TLSs \cite{AlOxmodel5}. Although there are difference between theoretical models, they are consistent in that H-TLSs have smaller $p$ than O-TLSs. 

The $\gamma-\mathrm{Al_{2}O_{3}}$ $D(p_z)$ has a single peak at approximately $p_z$ = 2.6 D, which can be sourced from H-TLS or two unresolved peaks of both H- and O-TLSs.
Although small in statistics, $\mathrm{a-Al}\mathrm{O_{x}}$ has a wider spread in $D(p_z)$ and two separate peaks. Furthermore, $\mathrm{a-Al}\mathrm{O_{x}}$ TLSs have 10 \% population with $p_z$ > 8.6 D, where $p_z$ = 8.6 D is the maximum in $\gamma-\mathrm{Al_{2}O_{3}}$. Using a comparison between two alumina datasets and the fact that O-based TLS is the larger dipole in DFT structures, we 
find a higher ratio of O-TLSs to H-TLSs in a-AlOx than $\gamma-\mathrm{Al_{2}O_{3}}$.

\section{Conclusion}

In summary, we have extracted the dipole moment
$p_{z}$ of hundreds of individual TLSs in nanoscale-thick films of
(polycrystalline) $\gamma-\mathrm{Al_{2}O_{3}}$ and (amorphous) $\mathrm{a-Al}\mathrm{O_{x}}$
alumina. We have used an Electrical-Bridge Quantum Defect Sensor (EBQuDS),
which we show is suitable to characterize number of TLSs as quantum
defects. Analysis of the measured histogram of $p_{z}$
reveals that polycrystalline alumina fits well to a single Gaussian
peak. From the material distribution (algebraically related to the
measured one), we obtain that the mean TLS moment of the polycrystalline
film is $p_{z}=$2.6$\pm0.3$ D (= 0.54$\pm0.05$ e$\mathring{A} $) and $\sigma$ =  1.6$\pm0.2$  D (=  0.33$\pm0.03$ e$\mathring{A} $). Furthermore, the material distribution
disagrees with the isotropic model commonly used in amorphous materials,
indicative of a preferred texture (orientation) of the polycrystalline
grains which host TLSs. 
On the other hand, we cannot conclude if amorphous alumina dipoles are isotropic or not because of missing TLSs extraction.

The ability to extract an accurate mean $p_z$ puts constraints on its defect type, and allows us to make first comparisons to new microscopic structures used in DFT calculations. 
The polycrystalline data show one dominant peak, and could be showing dominance of H-TLSs, or unresolved peaks of both H- and O-TLSs in the distribution. 
We find that $\mathrm{a-Al}\mathrm{O_{x}}$ has a larger mean $p_z$ = 4.6$\pm$0.5 D, which is consistent with previous amorphous alumina results. 
In contrast to $\gamma-\mathrm{Al_{2}O_{3}}$, the TLSs switch more rapidly, and our $p_z$ distribution in amorphous alumina yields a larger standard deviation (= 2.5$\pm$0.3 D) and two peaks in contrast to one. The moments above 8.6 D (10\% of the distribution) are larger than any TLS in polycrystalline alumina and agree only with calculations of delocalized O atoms. 
Due to this and other amorphous distribution features, we find that the ratio of O- to H-TLSs is higher in amorphous samples than in the polycrystalline ones. 
Because of its relative simplicity in distribution, alumina seems to be an important material for further JJ-barrier studies.

\section{Acknowledgements}

We acknowledge Lukas Radke, Hannes Rotzinger (both from Physikalisches Institut, Karlsruhe Institute of Technology, Germany) and Martin Weides (James Watt School of Engineering, University of Glasgow, UK) for support of discussions on amorphous alumina sample growth.

\setcounter{section}{0}
\appendix
\section{Material Density and Loss tangent}
\label{HtoD}
In this section, we derive the relationship between material TLS density,
$D(p_{z}),$ and the measured histogram, $H(p_{z})$. 
In the Standard Tunneling Model (STM) \cite{STM1,STM2}, the authors assumes that the density
of levels per unit volume and energy, $n(\Delta,\Delta_0)$, depends on tunneling energy $\Delta_0$ but is uniform in $\Delta$ giving 
\begin{equation}
n(\Delta,\Delta_0)\,d\Delta d\Delta_{0}\,=\,\frac{P_{0}}{\Delta_{0}}\,d\Delta d\Delta_{0},
\end{equation}
where $P_{0}$ is a constant in unit of $1\,/(\,J\,m^{3}$).
However, the model only considers a single moment |$\overrightarrow{p}$| = $p$. 
In our experiments, we notice the dipole moment in z axis is not uniform and we add dipole direction and magnitude dependence.
For a general case, we write
\begin{equation}
n(\Delta,\,\Delta_0,\overrightarrow{p})\,d\Delta d\Delta_{0} d^3p\,=\,\frac{P'_{0}}{\Delta_{0}}\,D(\overrightarrow{p})\,d\Delta d\Delta_{0} d^3p,
\end{equation}
where $D(\overrightarrow{p})$ is a generalized material TLS distribution and $P'_0$ is a new constant.
However, $D(\overrightarrow{p})$ depends on 3 Cartesian coordinates, and we only have measurement access to one component, $p_z$.
The full investigation of $D(\overrightarrow{p})$ is beyond the scope of this paper, but we assume that $D(\overrightarrow{p})$ is separable in $p_x$, $p_y$, and $p_z$.
Therefore, we consider the case
\begin{equation}
n(\Delta,\,\Delta_0,\,p_z)d\Delta d\Delta_{0} dp_z\,=\,\frac{D(p_z)}{\Delta_{0}}\,\,d\Delta d\Delta_{0} dp_z,\label{eq:Standard Dist}
\end{equation}
where $\intop D(p_{z})\,dp_{z}\,=\,P_{0}$ and $D(p_{z})$ is the material density mentioned in the main text.
In TLS spectroscopy, $V_{bias}$ is controlled such that $p_z$, $\Delta_0$, and $\Delta|_{V_{bias} = 0}$ can be extracted for individual TLSs and $\Delta\, =\, \Delta|_{V_{bias} = 0}\,+\,2\,p_{z}V_{bias}/l_{0}$, where $l_{0}$ is the thickness of dielectric.
Next, we change variables to include $V_{bias}$ through Jacobian transformation giving
\begin{equation}
d\Delta dp_{z}\,=\,dp_{z}dV_{bias}|\begin{array}{cc}
\frac{\partial\Delta}{\partial p_{z}} & \frac{\partial\Delta}{\partial V_{bias}}\\
\frac{\partial p_{z}}{\partial p_{z}} & \frac{\partial p_{z}}{\partial V_{bias}}
\end{array}|\,=(\frac{2\,p_{z}}{l_{0}})\,dp_{z}dV_{bias}\label{eq:changed basis}.
\end{equation}
$N_{tot}$ is the total number of observed TLSs from measurement histogram, $H(p_{zi})$,
\begin{equation}
N_{tot}\,=\,\sum_i\,H(p_{zi})\,\Delta p_{zi},
\end{equation}
where $p_{zi}$ is the center value and $\Delta p_{zi}$ is the bin width of the i-th bin.
$N_{tot}$ can also be written related to Eq. \ref{eq:Standard Dist} as
\begin{equation}
N_{tot}=\mathbb{V}\,\int n(\Delta,\,\Delta_0,\,p_z)d\Delta d\Delta_{0} dp_z.
\end{equation}
Substituting Eq. \ref{eq:changed basis} into the above equation, we have
\begin{equation}
N_{tot}\,=\mathbb{V}\,\int_{p_{min}}^{p_{max}}\,D(p_{z})\cdot(\frac{2\,p_{z}}{l_{0}})\,dp_{z}\,\int_{V_{1}}^{V_{2}}dV_{bias}\,\int\frac{d\Delta_{0}}{\Delta_{0}}.\label{eq: histogram}
\end{equation}
Similarly, we consider the i-th bin of $H(p_{zi})$ and the number of TLSs in this bin 
\begin{equation}
    N_i = H(p_{zi})\,\Delta p_{zi}
\end{equation}
In the case of small enough $\Delta p_{zi}$ and $\Delta_0$, we obtain
\begin{equation}
H(p_{zi})\,\Delta p_{zi}=\mathbb{V}\,(\frac{2\,p_{zi}}{l_{0}})\,D(p_{zi})\,\Delta p_{zi}\,\Delta V_{bias}\,\frac{\Delta f_{0}}{f_{0}}\label{eq: histogram and D relation}
\end{equation}
or
\begin{equation}
D(p_{z})\,=\frac{1}{\mathbb{V}}\,\frac{H(p_{z})}{2\,p_{z}}\,\frac{l_{0}}{\Delta V_{bias}}\,\frac{f_{0}}{\Delta f_{0}},\label{eq: D to H}
\end{equation}
where $\Delta f_{0}$ is the measurement frequency span. Thus, we prove that $D(p_{z})$ is not
proportional to measured histogram $H(p_{z})$, but $D(p_{z})\propto\frac{H(p_{z})}{p_{z}}$.

Next, we derive the loss tangent, $\mathrm{tan}\delta_{0}$, from TLS histogram.
Following Ref. \cite{Gaothesis}, loss due to TLSs can be described as
\[
\mathrm{tan}\delta\,=\,\int\frac{p_{z}^{2}}{\varepsilon}\,\frac{-\frac{1}{T_{2}}\,\mathrm{tanh}(\frac{E_{TLS}}{2k_{B}T})}{(T_2^{-2}+\Omega^{2}\frac{T_{1}}{T_{2}})+(\frac{E_{TLS}}{\hbar}-2\pi f_{0})^{2}}d^{3}n
\]

\begin{equation}
=\,\int D(p_{z})\frac{p_{z}^{2}}{\varepsilon}\,\frac{-\frac{1}{T_{2}}\,\mathrm{tanh}(\frac{E_{TLS}}{2k_{B}T})\,d\Delta\frac{d\Delta_{0}}{\Delta_{0}}}{(T_2^{-2}+\Omega^{2}\frac{T_{1}}{T_{2}})+(\frac{E_{TLS}}{\hbar}-2\pi f_{0})^{2}}dp_{z},
\end{equation}
where $\varepsilon$ is the permittivity constant, $\Omega$ is the
Rabi frequency, and $T_{1}$ ($T_{2}$) is TLS relaxation (decoherence)
time. The derivation of a similar integral has been presented in reference
\cite{Gaothesis}. In the case when $\Omega^2T_{1}T_2$ is much smaller than 1, 
\begin{equation}
\mathrm{tan}\delta\,=\,\frac{\pi}{\varepsilon}\int D(p_{z})p_{z}^{2}dp_{z}.
\end{equation}
Next, we estimate the loss tangent of $\gamma-\mathrm{Al_{2}O_{3}}$.
By replacing $D(p_{z})$ from Eq. \ref{eq: D to H}, we get
\[
\mathrm{tan}\delta=\frac{\pi}{\varepsilon}\,\frac{1}{\mathbb{V}}\,\int\frac{H(p_{z})}{2\,p_{z}}\,\frac{l_{0}}{\Delta V_{bias}}\,\frac{f_{0}}{\Delta f_{0}}\,p_{z}^{2}\,dp_{z}
\]

\[
=\,\frac{\pi}{2\,\varepsilon}\,\frac{l_{0}}{\mathbb{V\,}\Delta V_{bias}}\,\frac{f_{0}}{\Delta f_{0}}\,\int H(p_{z})\,p_{z}\,dp_{z}
\]

\[
\approx\,\frac{\pi}{2\,\varepsilon}\,\frac{l_{0}}{\mathbb{V\,}\Delta V_{bias}}\,\frac{f_{0}}{\Delta f_{0}}\,\sum H(p_{z})p_{z}\Delta p_{z}
\]

\begin{equation}
=\frac{\pi}{2\,\varepsilon}\,\frac{l_{0}}{\mathbb{V\,}\Delta V_{bias}}\,\frac{f_{0}}{\Delta f_{0}}\,\sum_{i}^{394}p_{zi}\,=\,1.4(1) \times 10^{-3}.
\end{equation}
The function is factor of $\pi$ larger than the equation (S6) in Ref. \cite{BahmanDCbiasTLS}. Notice that the loss tangent from the TLS histogram is similar to the bulk resonator loss tangent, $\mathrm{tan}\delta_{0}\,=\,1.5\times 10^{-3}$, reported
in main text. Last, we estimate
the material constant in the same film
\[
P_{0}\,=\,\int D(p_{z})dp_{z}
\]

\begin{equation}
\approx\,\sum_i\frac{1}{\mathbb{V}}\,\frac{1}{p_{zi}}\,\frac{l_{0}}{2\,\Delta V_{bias}}\,\frac{f_{0}}{\Delta f_{0}}\,=\,1.0(1)\times10^{44}(J^{-1}m^{-3}).
\end{equation}

In contrast to $\gamma-\mathrm{Al_{2}O_{3}}$, the loss tangent of amorphous alumina is 3.2(3) $\times 10^{-4}$ which is few times smaller than the measured amorphous alumina loss tangent (=9.8(9)$\times 10^{-4}$). It is expected because we do not fit those TLSs with uncompleted hyperbola curves.
\section{Internal Quality Factor Fitting And Bias Filtering}

\begin{figure}
\centering
\includegraphics[width=8.5cm]{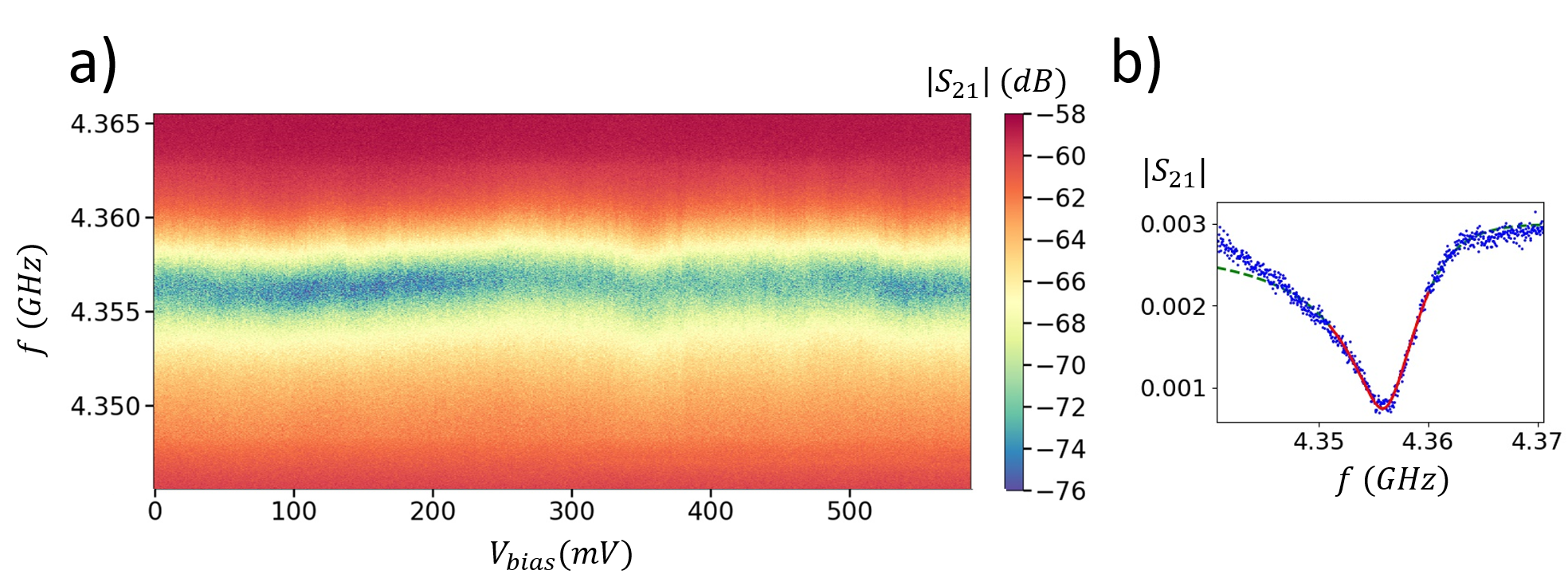}\caption{(a) Spectroscopy of DC bias sweep on $\gamma-\mathrm{Al_{2}O_{3}}$ with voltage noise. (b) One example of $|S_{21}|$ from (a) without proper filtering. Voltage noise obscures TLSs within the spectrum and $Q_i = 1600$.\label{Fig bad filtering}}
\end{figure}

\begin{figure}
    \centering
    \includegraphics[width=8.6cm]{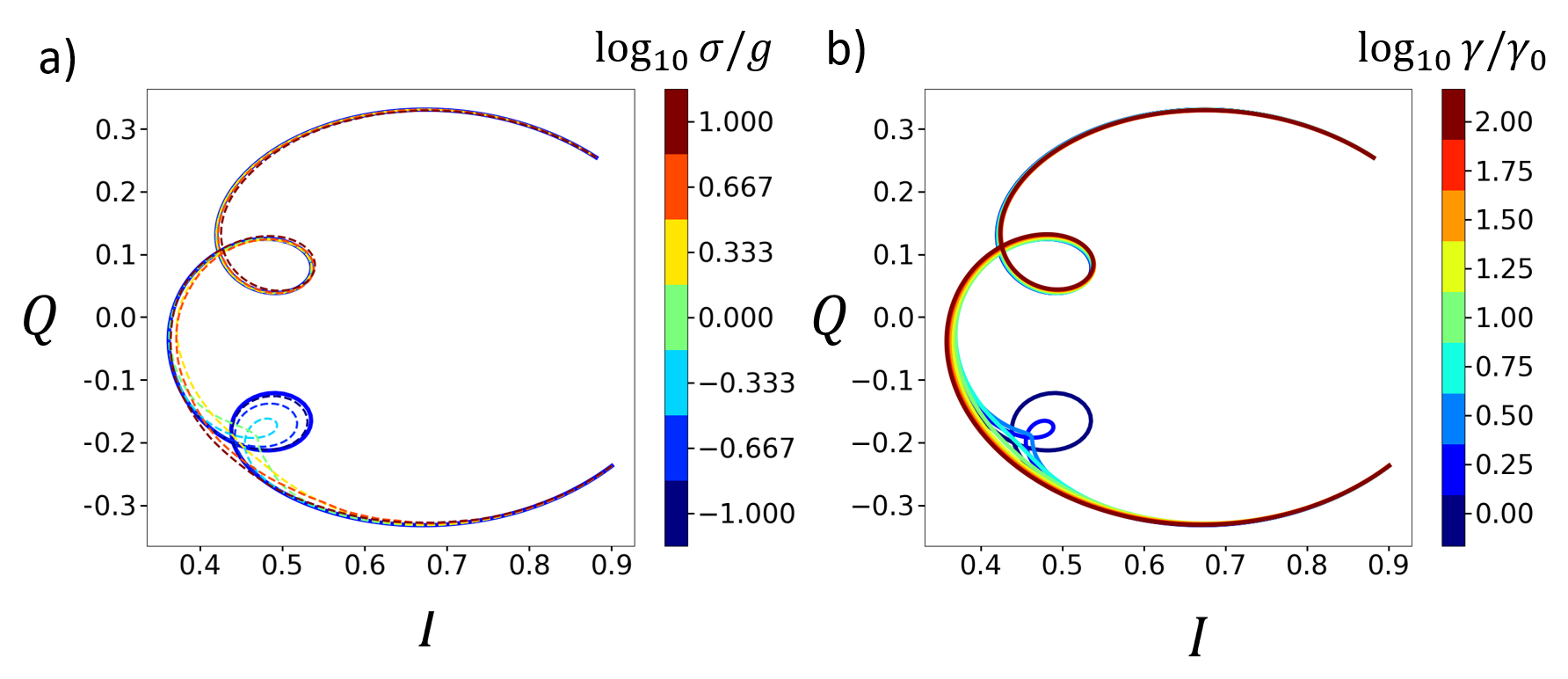}
    \caption{Simulation of Eq. \ref{eq: broaden S21} with resonator parameters $\kappa_{c},\,\gamma_{c}$ = 4 MHz, $f_c$ = 4 GHz, and two TLSs having $g$ = 0.7 MHz, $\gamma_0$ = 0.5 MHz. Only one of them is affected by voltage noise (lower one). (a) The solid blue line shows resonator coupled to two TLSs and both have no voltage noise. The colored dashed lines show the enhancement of $\sigma$ from 0.1 $\times \,g$ to 10 $\times \,g$ on the lower TLS. (b) Simulation of the lower TLS under enhanced decoherence rate. The color lines indicate the behavior of the lower TLS are similar to enhancement of $\sigma$ in (a).}
    \label{TLSwithnoise}
\end{figure}

In this section, we discuss the effect of bias line filtering and TLS noise on resonator data fittings. Filtering noise in the bias line is essential to study the
individual TLSs in both film types. We performed a control
experiment with additional bias-line noise. We start from a
setup where the bias line has only a low-pass copper powder
filter and a 12GHz K\&L filter. Fig. \ref{Fig bad filtering} (a) shows measurements of $\gamma-\mathrm{Al_{2}O_{3}}$ TLS spectroscopy with
low frequency noise and Fig. \ref{Fig bad filtering} (b) shows one |$S_{21}$| at fixed bias. There is no observation of any individual TLS. Surprisingly, the fit gives an internal quality factor, $\mathrm{Q_{i,noise}}$ = 1600, which is higher than $\mathrm{Q_i}$ = 680 reported in the main text. It is believed that without proper noise filtering, the bias voltage noise strongly affects the visibility of the TLSs. The change of TLS frequency due to voltage noise is 
\begin{equation}
\delta \omega_{TLS}\,=\,\frac{\Delta}{\hbar E_{TLS}}\delta\Delta\,=\,\frac{\Delta}{\hbar E_{TLS}}\times2p_{z}\delta V_{ex}/l_0. 
\end{equation}
In the case of multiple TLSs coupled to a resonator, the transmission rate \cite{BahmanThesis}
\begin{equation}
S_{21}(\omega)\,=\,1-\frac{\kappa_{c}/2}{\frac{\kappa_{c}+\gamma_{c}}{2}+i\,(\omega-\omega_{c})+\sum\frac{g_{i}^{2}}{\gamma_{i}/2\,+\,i\,(\omega-\omega_{i})}},\label{eq: TLS in S21}
\end{equation}
where $\kappa_{c}\,$is the cavity's decay rate to the transmission
line, $\gamma_{c}$ is the cavity decay rate to the environment when no
TLSs exist, $\omega_{c}$ is the cavity resonance, $g_{i}$ is the
coupling strength of each TLS, $\gamma_{i}$ is the decoherence rate, and
$\omega_{i}$ is the resonant frequency of the i-th TLS.
Inspired by Ref. \cite{fluctuations}, we assume that every TLS experiences a Gaussian voltage noise and we rewrite transmission rate as 
\begin{equation}
S_{21}\,=\,1-\frac{\kappa_{c}}{2}\,\dotsintop\frac{\prod_i\sqrt{\frac{1}{2\pi\sigma_{i}}}\,exp[\frac{-\delta\omega_{i}^{2}}{2\,\sigma_{i}}]\,d\delta\omega_{i}...d\delta\omega_{n}}{\frac{\kappa_{c} +\gamma_{c} }{2}+i\Delta_c+\sum\frac{g_{i}^{2}}{\frac{\gamma_{i} }{2}+i(\Delta_i+\delta\omega_{i})}},\label{eq: broaden S21}
\end{equation}
where $\Delta_c = \omega - \omega_c$ and $\Delta_i = \omega - \omega_i$yy
The standard deviation, $\sigma_{i}$, of the i-th TLS depends on $\delta V_{ex}$
and $p_{z}$ such that the sensitivity to voltage noise is described as an averaged effect. The enhancement of $\sigma_{i}$ exterminates the effect of the i-th TLS.
A simulation of different $\sigma_{i}$ is shown in Fig. \ref{TLSwithnoise} (a). Two TLSs are coupled to the resonator. The upper TLS in the plot is not affected by voltage noise and the lower TLS in the plot has $g$, $\gamma_0$, and a random frequency shift with deviation, $\sigma$. Simulations show that the lower TLS has smaller and smaller effect on the resonator while voltage noise (or $\sigma$) increases. The phenomena of the voltage noise is qualitatively similar to the increasing decoherence $\gamma$ of TLS as shown in Fig. \ref{TLSwithnoise} (b).

\section{Fitting of the averaged $S_{21}$ of $\mathrm{a-AlO_{x}}$}
\label{aAlOxfit}
Due to the low Q (<250) in $\mathrm{a-AlO_{x}}$ resonator, the resonator's transmission data is easily perturbed by frequency-dependent background.
For example, there will be (1) spurious modes: existing in the input-output cables, the sample box or on the chip and (2) electrical components: amplifiers, circulators, or attenuation.
One obvious result is that the off-resonant part of IQ plot deviates from the circle fitting curve where |$S_{21}$| is not constant. 
Thus, we modified the fitting function to better extract resonator's parameters. 
We consider the background as multiple low-Q modes, which have no direct coupling with each others and our resonator and the their transmission rate 
\begin{equation}
    S_{21,low Q} = \prod_{k=1}^{N} (1 - \frac{Q_k/Q_{e,k}}{1-2\,i\,Q_k\,(\omega-\omega_k)/\omega_k}),
\end{equation} 
where $Q_k, Q_{e,k}$ and $\omega_k$ are total quality factor, external quality factor and resonant frequency, respectively.
In general, in a small range around resonant frequency $\omega_0$, the background can be simplified to $(1 + (a + i b)(\omega-\omega_0))$, where a and b is a constant.
In the case that the $Q$ of the nearest mode to the measured resonator is in the order of 10 and $Q_e/Q_i \lesssim 1$ and each mode separates few hundreds of MHz, $S_{21,low Q}$ can be expanded around $\omega_0$ to the first order of $\omega$.
As a result, the final fitting function is
\begin{equation}
    S_{21} = C(1 + (a + i b)(\omega-\omega_0))e^{i\theta}(1 - \frac{Q/Q_e e^{i\phi}}{1 + 2 i \frac{Q}{\omega_0}(\omega-\omega_0)}).
\end{equation}

\begin{figure}
    \centering
    \includegraphics[width=8.6cm]{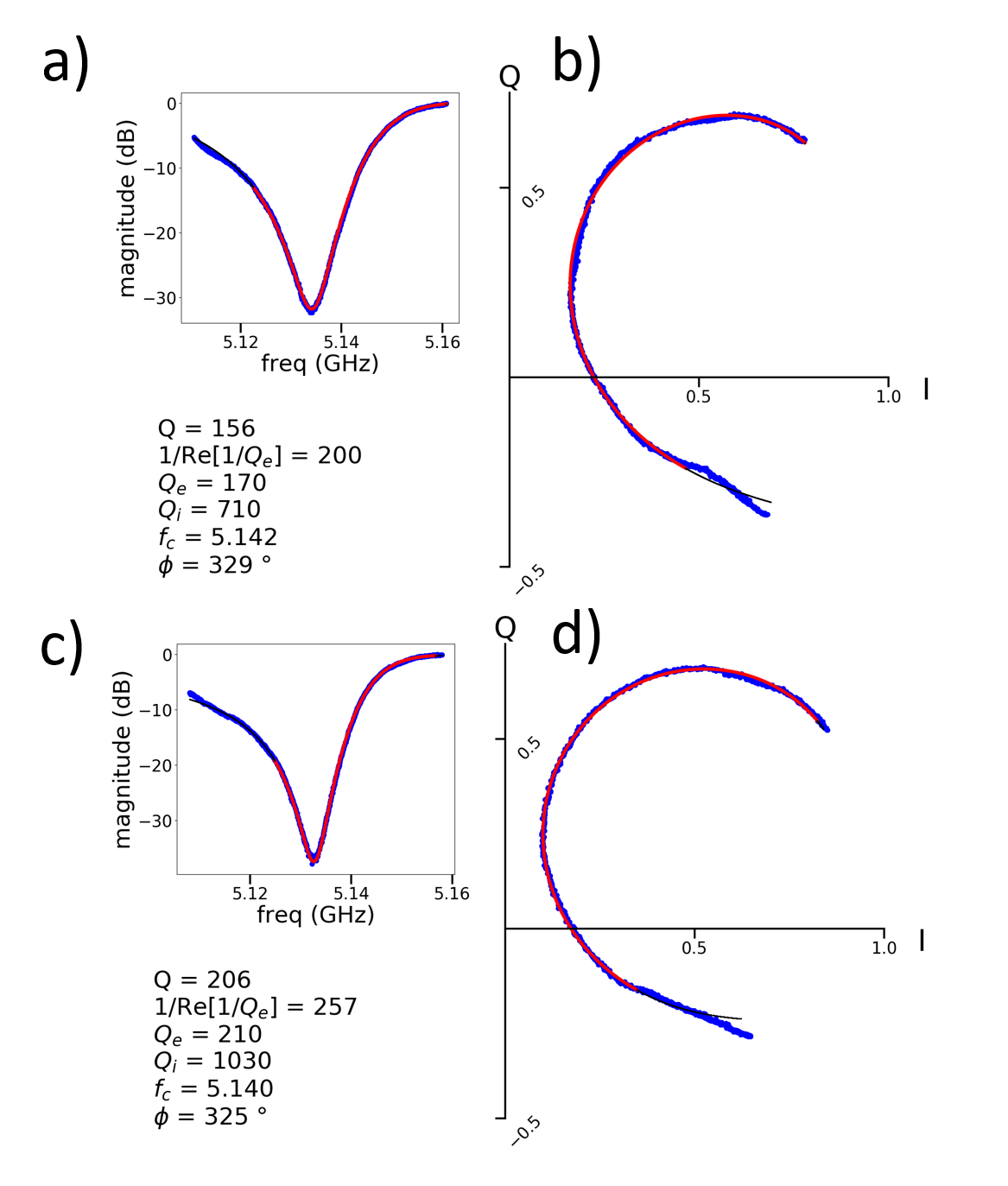}
    \caption{The normalized $S_{21,avg}$ of TLS spectrum of $\mathrm{a-AlO_{x}}$, which is similar to the black line in Fig. 1 (b). Unlike Fig 1 (b) in the main text, we add an frequency dependent on I and Q for a better fit. The red line represents the fitting range and the black line is the extension of fit to the measurement frequency range. Panel (a) and (b): the fitting result shows $Q_i$ = 710 and the dataset is from Fig. \ref{missedTLSspectrum} below. Panel (c) and (d): the dataset is shown in the supplementary. The fitting result shows $Q_i$ = 1030. }
    \label{average S21}
\end{figure}

Fig. \ref{average S21} shows two averaged $S_{21}$, which are normalized to 1, from two different cooldowns:
(a) and (b) are from the dataset of Fig. \ref{missedTLSspectrum}
; (c) and (d) are from the dataset shown in supplementary \cite{supplementary}.
To our surprise, the $Q_i$ of $\mathrm{a-AlO_{x}}$ sample are distinct in different cooldowns and the fits vary from 710 to 1510. We do not understand this variance in amorphous sample and no significant $Q_i$ changes for our $\gamma-\mathrm{Al_{2}O_{3}}$ sample in different cooldowns. 

The new averaged of $Q_i$ is 1020 for amorphous alumina which is 1.5 times higher than $\gamma-\mathrm{Al_{2}O_{3}}$. However, the loss tangent is lower in amorphous alumina can probably because of the below two reasons. First, amorphous alumina can have smaller density of TLSs with higher average dipole moment and lower loss tangent in the end. Second, amorphous alumina has higher voltage noise which eventually reduces the coherent time of TLS and loss tangent. Due to the smaller film thickness and a larger portion of large dipole TLS in $\mathrm{a-AlO_{x}}$, TLSs in $\mathrm{a-AlO_{x}}$ suffer from larger voltage noise than those in $\gamma-\mathrm{Al_{2}O_{3}}$. One example of extreme cases of high voltage noise is shown in Fig. S1, where we see no TLS and $Q_i = 1600$ is larger than $Q_i = 680$ reported in the main text.

\section{Fitting of hyperbolas in DC sweep plot}\label{Montefit}
\begin{figure}
    \includegraphics[width=8.6cm]{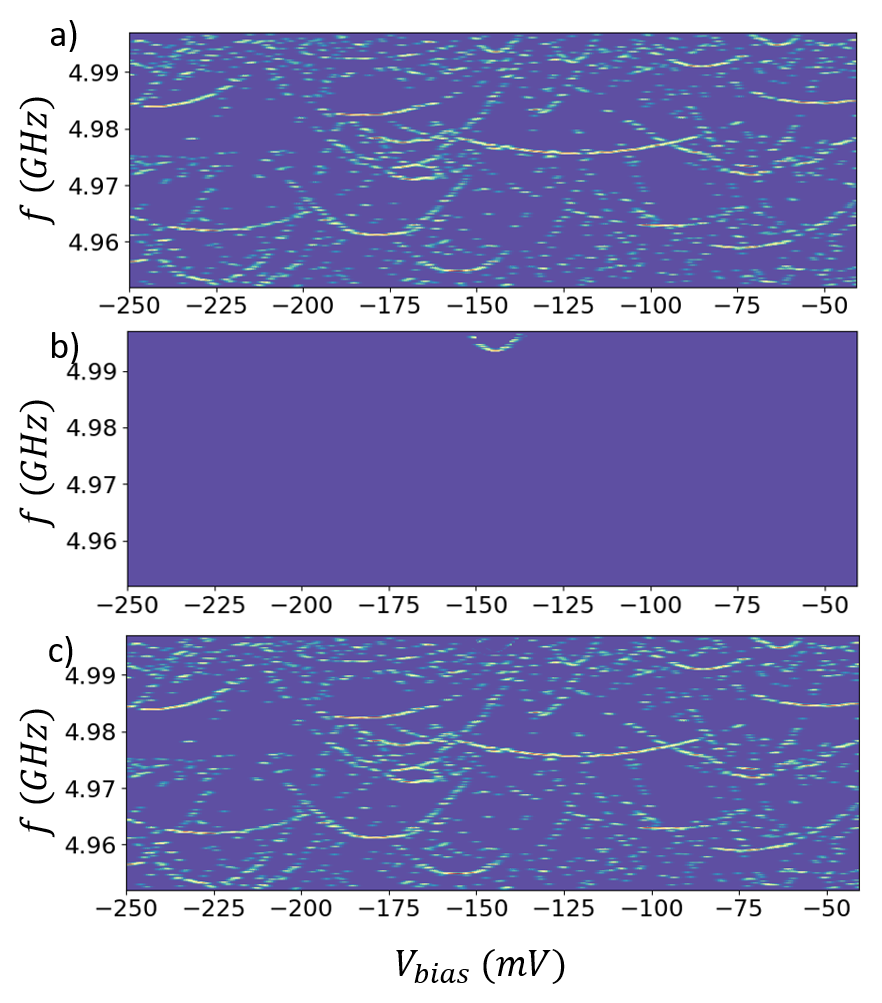}
    \centering
    \caption{(a) Plot of extracted minimum energies in Fig 2 of main text. (b)
The final result of first dipole fit. (c) Subtract the first dipole
fitting result from (a). \label{fig: dipole fitting}}
\end{figure}

This section is describing the procedure of fitting TLS dipole for Fig. 2 and 3 in
the main text. We search the local minimum (hybridized state) of each $\left|S_{21}\right|$ from Fig. 2 (a) and apply a Gaussian filter on the 2D plot of local minimum (see Fig. \ref{fig: dipole fitting} (a) for example). 
We obtain the initial guess values of $\Delta_{0}$, $V_{bias}$, and $p_{z}$ for all possible TLS candidates.
Due to the complexity of TLS spectrum, TLSs are fit separately in the order of their $\Delta_{0}$ from the largest to the smallest.
The first candidate TLS with the largest $\Delta_{0}$ is fitted by Monte Carlo method and the result is shown in Fig \ref{fig: dipole fitting} (b).
We assign those local minimum to first TLS and subtract them from Fig. \ref{fig: dipole fitting} (a) to get Fig. \ref{fig: dipole fitting} (c).
For next TLS, we use the local minimum plot from Fig. \ref{fig: dipole fitting} (c) and the next fitting will not be affected by the previous TLS result.
We repeat the process until all TLSs are fit and discard those TLSs not crossing their minimum energies.

\section{$\gamma-\mathrm{Al_{2}O_{3}}$ TLS histograms in two different resonators}
\label{tworesHis}
\begin{figure}
\centering
\includegraphics[width=8.6cm]{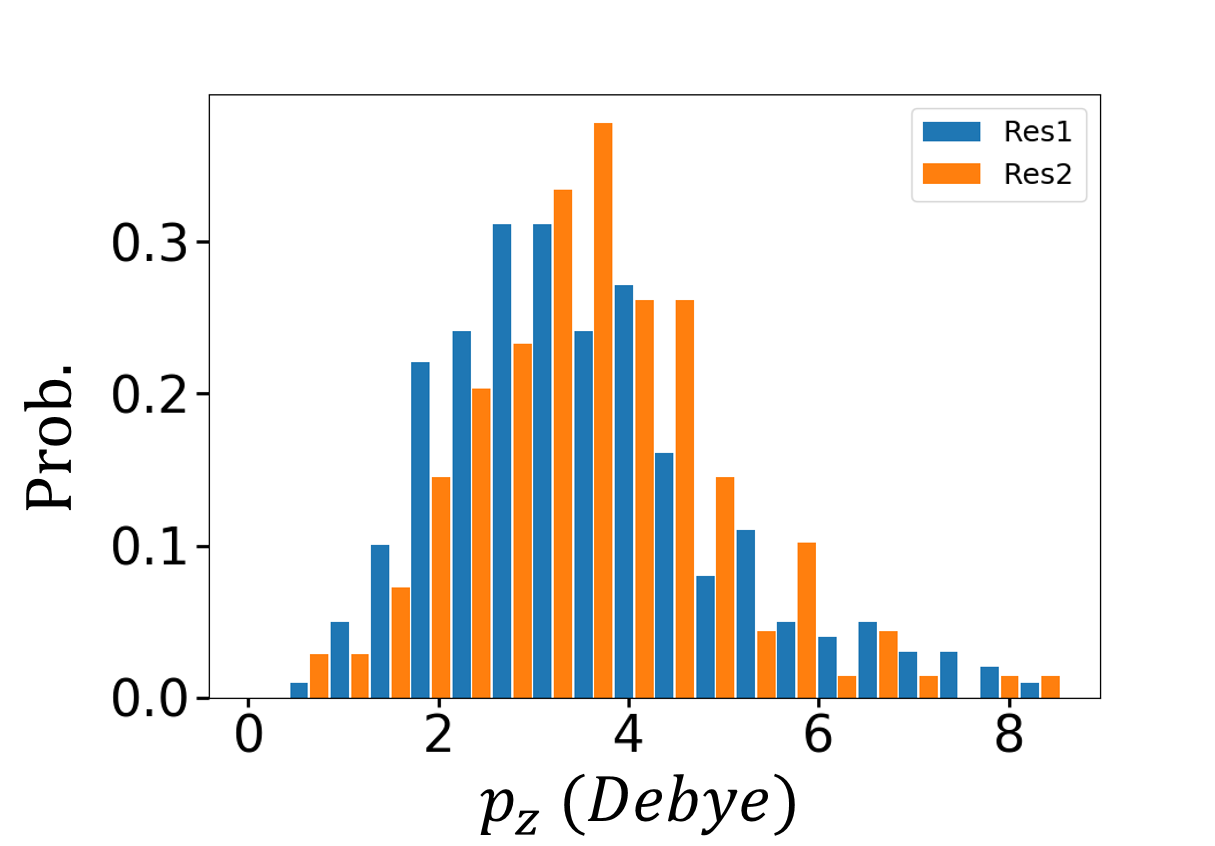}\caption{Histograms of dipole moments from two different $\mathrm{\mathrm{\gamma-Al}_{2}\mathrm{O}_{3}}$ resonators. }
\label{twoTLSdipolehistogram}
\end{figure}

\begin{figure}
\centering
\includegraphics[width=8.6cm]{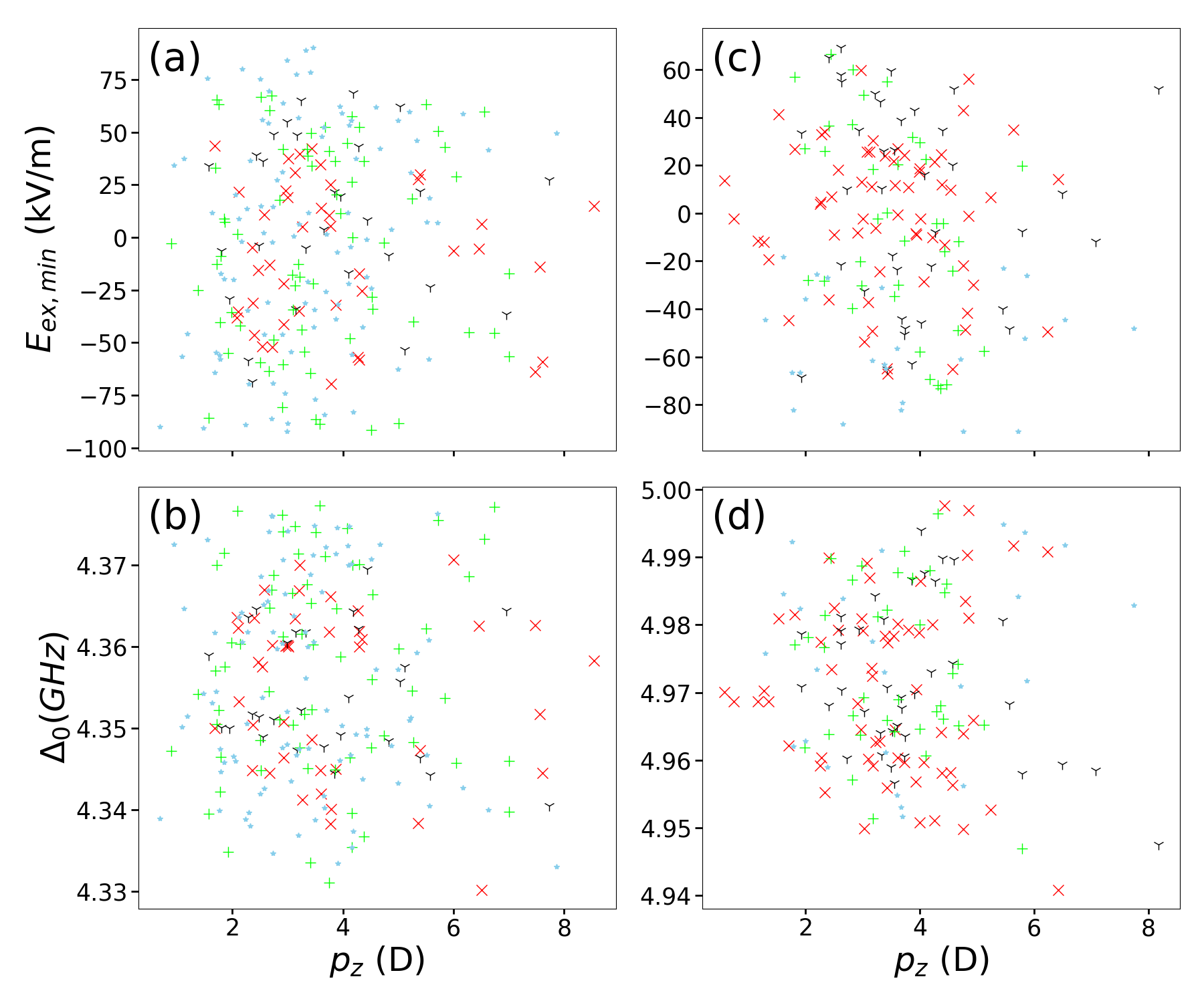}
\caption{Dipole moments $p_z$ vs $\Delta_0$ or the field of minimum TLS energy $E_{ex,min}$ in four cooldowns. Panel (a) and (b) is from the resonator of frequency $\approx 4.35$ GHz. Panel (c)and (d) is from the resonator of frequency $\approx 4.97 $GHz. Markers with different color represent four different cooldowns.}
\label{scatter_dipole}
\end{figure}

Fig. \ref{twoTLSdipolehistogram} and Table. \ref{tabletwoRes} show the dipole moment measured in two $\gamma-\mathrm{Al_{2}O_{3}}$ resonators. The mean
$p_z$ are 3.5(1) and 3.6(1), for Res1 with $f_0$ = 4.35 GHz and Res2 with $f_0$ = 4.95 GHz, respectively. Fig \ref{scatter_dipole} shows the dipoles measured in 4 different cooldowns vs either $\Delta_0$ or the external field $E_{ex,min}$, where TLSs meet their minimum energies. 

Amorphous materials rearrange their atom positions when their temperature is above glass transition temperature ($T_g$).
The thermal cycle to room temperature ($\leq T_g$) is likely not enough.
There are chances that TLSs are measured twice in two cooldowns.
However, we found the $p_z$ of different cooldowns scatter randomly in the plots (see Fig. \ref{scatter_dipole}). Theoretically, TLSs do not change their $\Delta_0$ without reaching $T_g$. However, we cannot tell if TLSs' $\Delta_0$ and $p_z$ have changed sufficiently to be considered as new TLSs. But from Fig. \ref{scatter_dipole} (b) and (d), most TLSs ($\gtrsim$ 85\%) have different $\Delta_0$ and $p_z$ in different cooldowns. Whether the thermal cycle is constructing a new set of TLSs is beyond the scope of this paper.

\begin{table}[!htbp]
\centering
\begin{tabular}{|l|l|l|}
\hline 
& mean $p_z$ (D)  & standard deviation (D)\\
\hline 
\hline 
Res1 (4.35 GHz)  & 3.55  & 1.47 \\
\hline 
Res2 (4.95 GHz)  & 3.64  & 1.3 \\
\hline 
\end{tabular}
\label{tabletwoRes}
\caption{Mean and standard deviation of extracted TLS $p_z$ for each $\gamma-\mathrm{Al_{2}O_{3}}$ resonator.}
\end{table}

\section{Fitting of TLS density by maximum likelihood estimation}

Here, we show the procedure of applying Fisher's maximum likelihood estimation
(MLE) on our statistics. For simplicity, we choose truncated normal distribution as our target function $f(p_{z};\,\mu,\,\sigma)$ to fit our material density $D(p_z)$, and
\begin{equation}
f(p_{z};\,\mu,\,\sigma)\,=\,C(\mu,\sigma)\,\frac{1}{\sigma\sqrt{2\pi}}exp(-\frac{(p_{z}\,-\,\mu)^{2}}{2\sigma^{2}})
\end{equation}
$\mathrm{for}\,p_{z}\in[0,\infty)$, where $\mu$ is the mean value, $\sigma$ is the standard deviation. The normalized constant
\begin{equation}
C(\mu,\sigma)\,=\,\frac{2}{1-\mathrm{erf}(-\mu/\sigma\sqrt{2})}
\end{equation}
depends only on $\mu$ and $\sigma$,
where erf(x) is error function defined as 
\begin{equation}
\mathrm{erf}(x)\,=\,\frac{2}{\sqrt{\pi}}\int_{0}^{x}\mathrm{exp}(-t^{2})dt.
\end{equation}
The likelihood function
\begin{equation}
L\,=\,\prod_{i}f(p_{zi};\,\mu,\,\sigma).
\end{equation}
The necessary conditions for the occurrence of a maximum (or a minimum) are
\begin{equation}
\frac{\partial ln(L)}{\partial\mu}\,=\,0,\,\frac{\partial ln(L)}{\partial\sigma}\,=\,0.
\end{equation}
As mentioned above in Sec. S-I, $D(p_{z})$ is not a directly measurement result, but $H(p_{z})$ is. Since there is a weighing factor $p_z$ transferring $D(p_{z})$ to $H(p_{z})$ , we have
\begin{equation}
H(p_{z};\,\mu,\,\sigma)\,=\,N_{tot}\,C_{1}(\mu,\sigma)p_{z}\,\mathrm{exp}(-\frac{(p_{z}\,-\,\mu)^{2}}{2\sigma^{2}})\label{measured TLS fitted function}.
\end{equation}
A new normalization constant is
\begin{equation}
C_{1}(\mu,\sigma)\,=\,\sigma^{2}\mathrm{exp}(-\frac{\mu^{2}}{2\sigma^{2}})+\,\mu\,\sigma\,\sqrt{\frac{\pi}{2}}\,(\,1\,-\mathrm{erf}(-\frac{\mu}{\sigma\sqrt{2}})),
\end{equation}
and $N_{tot}$ is the total observed TLS.

We also apply a gamma distribution function to fit the data, where
\begin{equation}
f(x,\alpha,\beta)\,=\,\frac{x^\alpha\,e^{-\beta\,x}}{\Gamma(\alpha)\beta^\alpha},
\end{equation}
and $\Gamma(\alpha)$ is a gamma function. The fitting result gives $\alpha$ = 5.15 and $\beta$ = 1.72 and mean = 2.99 D. However, we expect there is a distribution > 0 when dipole equals 0 so that gamma distribution is not suitable. 

MLE method applied to the amorphous alumina measured histogram does not give the main peak feature adequately in the material distribution $D(p_z)$. Thus, we do not use MLE to report averaged dipole in the amorphous data, but rather a calculated mean and standard deviation. The calculation method is described in the main text.
\begin{figure}
\centering
\includegraphics[width=8.6cm]{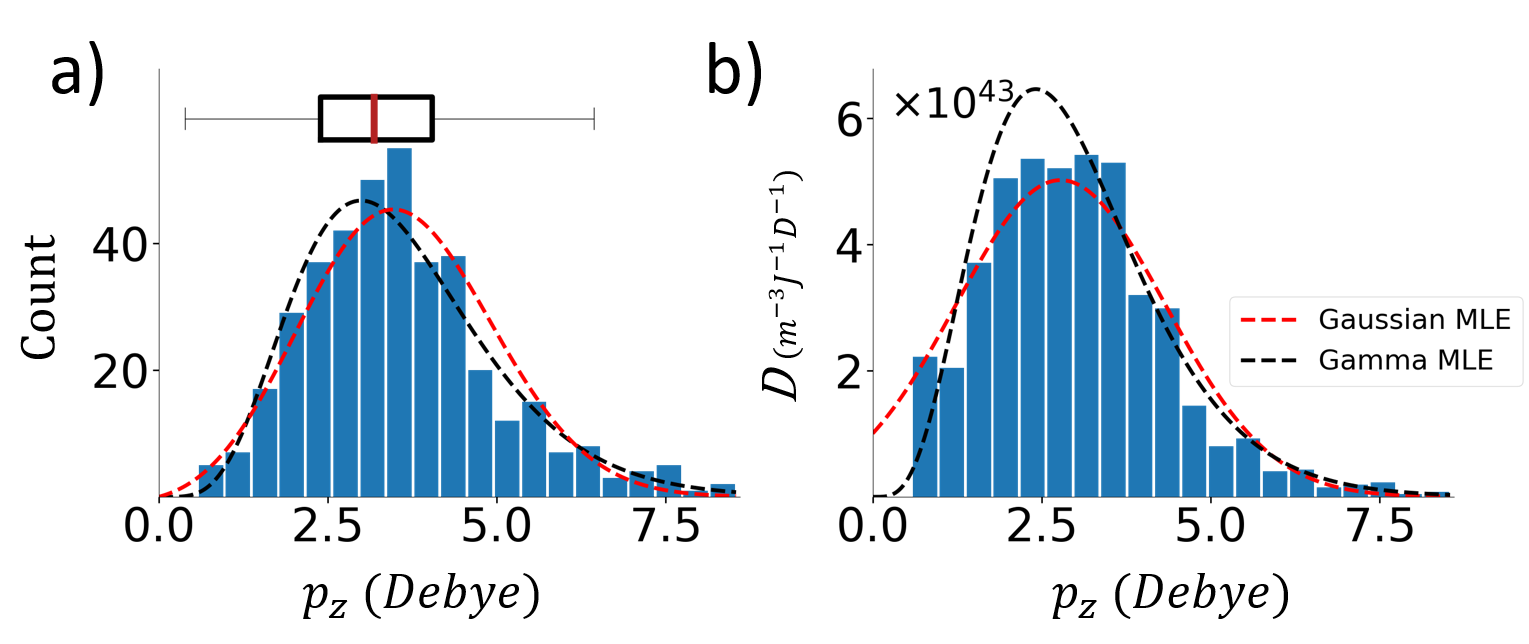}
\caption{Gamma distribution MLE analysis on $\gamma-\mathrm{Al_{2}O_{3}}$ data added to Fig. 2 (c) and (d) in the main text. Panel (a): measured distribution (b) Material distribution. The black dashed lines use the target function of Eq. S28 to fit the distribution. The red dashed lines are Gaussian function as the main text. The Gaussian function shows a better agreement to the data than Gamma distribution.}
\label{gamma_distribution}
\end{figure}

\section{Estimation of averaged dipole in $a-\mathrm{AlO_{x}}$ under TLS frequency noise}
\label{overestimation}
\begin{figure}
\centering
\includegraphics[width=8.6cm]{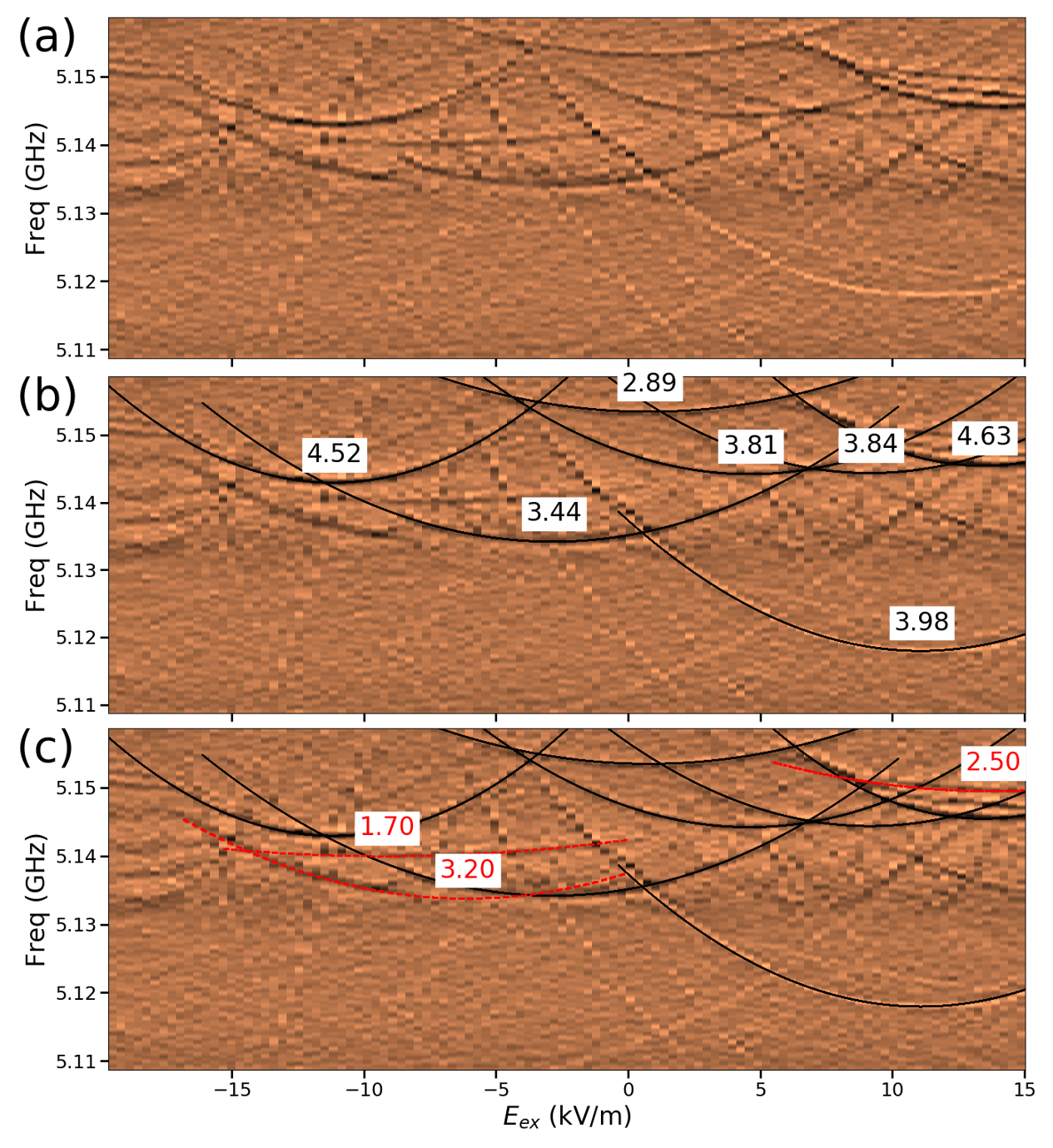}
\caption{One TLS spectroscopy of $a-\mathrm{AlO_{x}}$ TLS. (a) original data and (b) TLS fitting results and their dipole $p_z$ (D). (c) TLSs with incomplete hyperbolas in red and their potential $p_z$ (D)}
\label{missedTLSspectrum}
\end{figure}
\begin{figure}
\vspace*{\floatsep}
\centering
\includegraphics[width=8.6cm]{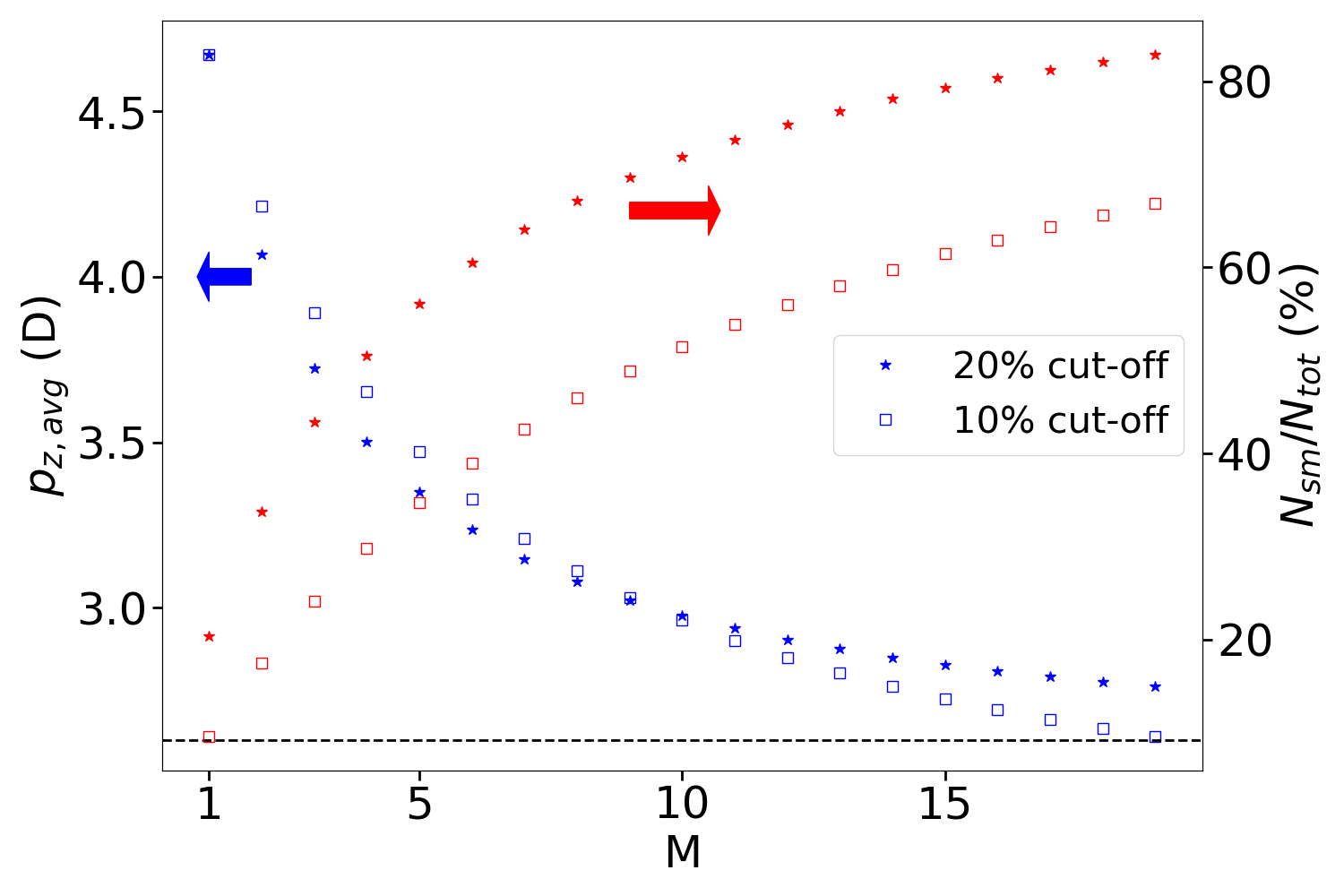}
\caption{Computed average dipole for different conditions of plausible missing TLSs in the amorphous film. We assume 10th (2.8 Debye) or 20th (3.5 Debye) percentile in original data as small $p_z$ TLS in square or in star marks respectively. Small $p_z$ TLSs in the distribution are multiplied by the multiplication factor M to simulate possible missing TLSs. The original data is when M = 1. The black dashed line represents the average dipole in $\gamma-\mathrm{Al_{2}O_{3}}$. We take M=5 as the most possible case for our data, implying $p_{z,avg}$ > 3.3 D in amorphous alumina film. }
\label{averagenewdipole}
\end{figure}
Because of the unstable TLS during voltage bias, and resonator noise level, we are prone to measure the larger dipole moments than the smaller ones especially in amorphous sample.
Those TLSs without crossing their minimum are not counted in statistics due to the high uncertainty.
We show few examples in Fig. \ref{missedTLSspectrum} (c), where preliminary manual fits are shown in dashed line.
A closer look, one can see some other potential small dipole moment TLSs at the edges of the figure. 

Here we estimate a case when we miss M-1 out of M small dipole moment by multiplying the small $p_z$ TLS distribution. First, we define the small $p_z$ TLSs as the 10th or 20th percentile, which equals 2.8 and 3.5 Debye respectively. The dipole smaller than the above value are multiplied M times in counts for a new distribution. The resulting new average dipole $p_z$ and ratio of small $p_z$ TLS number to the total number are plotted in Fig. \ref{averagenewdipole}. 
Suppose an extreme case with 8 missing TLS and 2 fitted TLS with $p_z$ < 3.5 D, it means 4 out of 5 are missed.
This implies multiple factor M of 5 and the new average is at least 3.3 Debye.
This is still larger than $\bar{p_z}$ = 2.6 Debye in $\gamma-\mathrm{Al_{2}O_{3}}$. Now, we turn to the isotropy of dipole orientation. There are missing TLS extractions, especially of those small $p_z$ TLS.
The portion of small $p_z$ TLS could be underestimated. 
As a result, if we add those small $p_z$ back, the material distribution $D(p_z)$ could be monotonic decreasing as expected in the standard model. 
More studies are needed to understand the small dipole TLSs in amorphous phase.


\bibliography{Al2O3_reference}

\end{document}